\documentclass[%
 preprint,
 superscriptaddress,
 amsmath,amssymb,
 aps,
 showkeys,
 titlepage,
 endfloats*.   %move all floats to the end of the document and put each on a separate page % superscript references
]{revtex4-2}

\newcommand{\code}[1]{\texttt{#1}}
\usepackage[version=4]{mhchem}
\usepackage{chemformula}
\usepackage{graphicx}% Include figure files
\usepackage{dcolumn}% Align table columns on decimal point
\usepackage{bm}% bold math
\usepackage[mathlines]{lineno}% Enable numbering of text and display math
\usepackage{siunitx}
\usepackage[colorlinks = true,
            linkcolor = blue,
            urlcolor  = blue,
            citecolor = red,
            anchorcolor = blue]{hyperref}%make cross-references clickable
\begin{document}

%\preprint{APS/123-QED}

\title{Lattice Dynamics and Thermal Transport in Semiconductors with Anti-bonding Valence Bands} 

\author{Jiaoyue Yuan}
\thanks{These authors contributed equally to this work.}
\affiliation{Department of Mechanical Engineering, University of California, Santa Barbara, CA 93106, USA}
\affiliation{Department of Physics, University of California, Santa Barbara, CA 93106, USA}

\author{Yubi Chen}
\thanks{These authors contributed equally to this work.}
\affiliation{Department of Mechanical Engineering, University of California, Santa Barbara, CA 93106, USA}
\affiliation{Department of Physics, University of California, Santa Barbara, CA 93106, USA}

\author{Bolin Liao}
\email{bliao@ucsb.edu} \affiliation{Department of Mechanical Engineering, University of California, Santa Barbara, CA 93106, USA}

\date{\today}

\begin{abstract}
Achieving high thermoelectric performance requires efficient manipulation of thermal conductivity and a fundamental understanding of the microscopic mechanisms of phonon transport in crystalline solids. 
One of the major challenges in thermal transport is achieving ultralow lattice thermal conductivity. 
In this study, we use the anti-bonding character of the highest-occupied valence band as an efficient descriptor for discovering new materials with an ultralow thermal conductivity. We first examined the relationship between anti-bonding valence bands and low lattice thermal conductivity in model systems \ce{PbTe} and \ce{CsPbBr3}. 
Then, we conducted a high-throughput search in the Materials Project database and identifed over 600 experimentally stable binary semiconductors with a strong anti-bonding character in their valence bands. 
From our candidate list, we conducted a comprehensive analysis of the chemical bonds and the thermal transport in the XS family, where X=K, Rb, and Cs are alkaline metals. 
These materials all exhibit ultralow thermal conductivities less than 1~W/(m K) at room temperature despite simple structures.
We attributed the ultralow thermal conductivity to the weakened bonds and increased phonon anharmonicity due to their anti-bonding valence bands. 
Our results provide chemical intuitions to understand lattice dynamics in crystals and open up a convenient venue towards searching for materials with an intrinsically low lattice thermal conductivity. 
\end{abstract}

                            
\maketitle

%\tableofcontents
% \renewcommand\linenumberfont{\normalfont\tiny}

% \linenumbers\relax % Commence numbering lines

\section{Introduction}\label{sec:intro}

 Crystalline solids with a low thermal conductivity have been extensively studied for applications such as thermoelectric materials~\cite{mao2018advances,chen2018manipulation,qian2021phonon,liao2015nanocomposites}, thermal insulation and thermal barrier coatings~\cite{clarke2005thermal}.
For example, to achieve highly efficient thermoelectric power generation or cooling, one has to overcome the challenge to minimize the lattice thermal conductivity $\kappa$ while maintaining the electrical conductivity $\sigma$ to maximize the thermoelectric figure of merit $ZT=\frac{\sigma S^2 T}{\kappa}$, wherein $S$ is the Seebeck coefficient and $T$ is the temperature. %mobility
 A high figure of merit $ZT$ in crystalline solids requires that the electron flow remains unimpeded while phonons get heavily scattered, a scenario known as the ``phonon-glass, electron-crystal'' (PGEC)~\cite{nolas2001semiconductor,beekman2015better,takabatake2014phonon,keppens2000does}.
 Some of the most successful techniques developed for the manipulation of phonon transport include introducing defects~\cite{snyder2008complex,minnich2009bulk}, nano/microstructural modifications using grain-boundary engineering~\cite{biswas2012high,poudel2008high} and interfaces~\cite{ravichandran2014crossover}. However, these extrinsic approaches all introduce defects and/or interfaces that can also interfere with electron transport~\cite{liao2015nanocomposites}. In this light, crystalline materials with an intrinsic ultralow thermal conductivity are particularly desirable.
 In previous studies, intrinsically low thermal conductivities have been achieved by exploring phonon scattering mechanisms originating from chemical bonding and structural aspects~\cite{sun2020high,ma2013glass,zhao2014ultralow,lu2013high,he2022accelerated}.
These include layered structures~\cite{chiritescu2007ultralow,costescu2004ultra}, liquid-like sublattices~\cite{liu2012copper,roychowdhury2018soft,li2018liquid}, local structural distortions~\cite{xie2020all}, ferroelectric-instability-induced phonon softening~\cite{delaire2011giant,sarkar2020ferroelectric}, rattling phonon modes~\cite{tritt2001recent,tadano2015impact,lin2016concerted,he2016ultralow}, and anharmonic lattice vibrations originating from electron lone pairs~\cite{nielsen2013lone,skoug2011role,dutta2019bonding}. 

At a microscopic level, the intrinsic lattice thermal conductivity of crystalline materials is largely controlled by the chemical bonding strength, which significantly impacts the velocity of the heat-carrying acoustic phonons. 
This explains the much lower thermal conductivity in materials with dominant ionic or van der Waals bonds than those with dominant covalent bonds~\cite{nolas2001thermoelectrics}. 
However, ionic and van der Waals bonds tend to have limited overlap between neighboring electronic orbitals, leading to weak electronic dispersion and low electron mobility. 
To balance the impact of the chemical bonds on the transport properties of both phonons and electrons, special types of covalent bonds with a mixed ionic character have been pursued. 
One prominent example is the abnormally low intrinsic lattice thermal conductivity of the IV-VI semiconductor family that hosts many of the best known thermoelectric materials, including PbTe, SnSe and GeTe~\cite{chen2017lattice,zhao2016ultrahigh,li2018low,liu2020high}. 
Despite their simple and high-symmetry crystal structures, these materials exhibit much lower thermal conductivity than other materials with similar atomic masses~\cite{lee2014resonant}. 
One common feature underlying their low lattice thermal conductivity is the presence of soft optical phonon modes that can strongly scatter heat-carrying low-frequency acoustic phonons, which originate from the long-ranged interatomic interactions due to resonant chemical bonds~\cite{lee2014resonant,yue2023strong}. 
It is also understood that the lone electron pair associated with the group IV element in these materials plays an important role in their unusual bonding structure~\cite{nielsen2013lone,zhang2023dynamic,wan2019strong}. 
Due to their mixed ionic-covalent bonding character, the electron mobility is not compromised in these materials, making them promising platforms to realize PGEC. 
Another emerging example with an ultralow lattice thermal conductivity is the lead halide perovskites (LHPs), which have attracted significant recent attention for photovoltaic applications~\cite{kim2020high}. 
Both fully inorganic and inorganic-organic hybrid versions of these materials have a thermal conductivity below 1 W/(m K) at room temperature, which has been attributed to soft chemical bonds\cite{ma2019supercompliant}. 
It has been increasingly recognized that the mixed ionic-covalent character of the chemical bonds in LHPs, likely also linked to the lone electron pair in Pb$^{2+}$ ions~\cite{fabini2020underappreciated}, is not only responsible for their strong lattice anharmonicity, low lattice thermal conductivity and facile ion migration~\cite{kim2023mapping}, but also their extraordinary optoelectronic and electron transport properties. 
Given the detailed understanding of the unusual lattice properties of these materials, the remaining challenge is to efficiently identify other materials with similar characters that can be promising candidates for applications requiring an intrinsically low thermal conductivity.

Indeed, the quests for new materials with extremely low thermal conductivity have been abundant. 
Previous studies have utilized high-throughput screening techniques to identify potential candidates based on physical characteristics such as large atomic mass~\cite{wang2020cu3erte3}, structural information of rattling atoms~\cite{li2022high}, and the combination of machine-learning algorithms and automatic \textit{ab initio} calculations~\cite{carrete2014finding,rodriguez2023million}. 
However, a comprehensive search focusing on the chemical bonding character and its impact on lattice dynamics has not been carried out due to the lack of an effective descriptor. In this work, we focus on a chemical bonding signature: highest-occupied valence bands with a strong anti-bonding character in a semiconductor. Recent studies have suggested that anti-bonding chemical bonds are closely related to ultralow thermal conductivities in a range of materials~\cite{yuan2022antibonding,das2023strong,zhang2022antibonding}. The advantage of using the anti-bonding character of the highest-occupied valence band as a descriptor is that it can be efficiently analyzed using 
 the crystal orbital Hamilton populations (COHP) method~\cite{dronskowski1993crystal, deringer2011crystal}, which only requires ground-state density functional theory (DFT) calculations.
This method can be applied to any inorganic crystal structure and requires only basic structural and compositional information as input and minimal time and computing resources. The low computational cost makes it a suitable method to screen materials in databases to identify new semiconductors with an ultralow lattice thermal conductivity. In this paper, we first show that the strong anti-bonding valence bands (ABVBs) underlie the unusual lattice dynamics in known example materials with an ultralow thermal conductivity, such as \ce{PbTe} and \ce{CsPbBr3}. In particular, the highest occupied valence bands with a strong anti-bonding character not only lead to weakened chemical bonds, but also give rise to soft optical phonons. Then, we conduct a high-throughput screen within the Materials Project database~\cite{jain2013commentary} to search for binary semiconductors with a strong ABVB. 
As a result, we identified more than 600 binary semiconductors with an anti-bonding valence band from a pool of over 6,000 candidates, allowing for the possibility of realizing ultralow intrinsic thermal conductivities. 
Out of the identified candidates, we highlight the XS (X=Na, K, Rb, and Cs) semiconductor family, where the sulfur ions are in an unusual valence state (S$^{-}$). We found that they all exhibit ultralow thermal conductivities despite having simple crystal structures, which can be attributed to their strongly ABVBs. Our findings suggest that a highest-occupied valence band with a strong anti-bonding character is indicative of impeded thermal transport and our high-throughput screening strategy also offers a novel approach for identifying materials with an intrinsically low thermal conductivity.

\section{Methods}\label{sec:method}

Our density functional theory (DFT) electronic structure calculations were performed using the Vienna Ab initio Simulation Package (\code{VASP})~\cite{kresse1996vasp1,kresse1996vasp2} with
the projector augmented wave (PAW) method~\cite{blochl1994projector} and the Perdew-Burke-Ernzerhof form of the generalized gradient approximation (PBE-GGA) of the exchange-correlation functional~\cite{perdew1996generalized}.
The valence wavefunctions were expanded on a plane-wave basis with a 400\,eV energy cutoff for all the materials studied in this work. 
The spin polarization and spin-orbit interaction were explicitly taken into account. The energy and force convergence criteria were set to be 1 $\times$ 10$^{-7}$ and 0.01 eV/\AA, respectively. 
The $\Gamma$-centered $n \times n \times n$ (n = 1, 2, 3, or 4) \textbf{k}-point grids were used to sample the first Brillouin zone depending on the unit cell size during high-throughput material screening. For selected materials studied in this paper, further \textbf{k}-point convergence was tested to make sure the lattice parameters and forces were converged. Particularly, the bonding nature associated with various electron energy bands were analyzed using the COHP method~\cite{dronskowski1993crystal,deringer2011crystal}. 
The COHP method partitions the energy of the band structure into interactions between pairs of atomic orbitals between adjacent atoms.
It is a bond-weighted measure of the electronic density of states (DOS) and provides a quantitative measure of the bonding and anti-bonding contributions to the band energy.
Importantly for this work, the sign of the COHP differs for bonds with bonding or anti-bonding nature: a positive (negative) sign corresponds to anti-bonding (bonding) interactions. 
By convention, COHP diagrams plot the negative value (-pCOHP) such that bonding (anti-bonding) states on the right (left) of the axis can be easily visualized.
We further quantify the degree of anti-bonding using the integrated area under the COHP curves with respect to the electron band energy. 

The phonon dispersion relations were obtained by conducting finite-displacement force calculations~\cite{broido2007intrinsic}, from which the harmonic interatomic force constants (IFCs) were extracted.
Then the dynamic matrices were constructed and diagonalized using the Phonopy package to generate phonon eigenfrequencies~\cite{togo2015first}.
Convergence of the phonon dispersions with respect to the supercell size and the \textbf{k}-grid sampling were tested in materials selected for a detailed study.
The phonon dispersion calculation of polar materials also included the non-analytic polar corrections. We calculated the lattice thermal conductivity ${\kappa}_\text{ph}$ by iteratively solving the phonon Boltzmann transport equation using the ShengBTE package~\cite{li2014shengbte}. Using the finite displacement method, we computed the anharmonic 3rd-order IFCs. The 3rd-order finite displacement calculation employed the identical supercell size and \textbf{k}-grid sampling as the 2nd-order IFCs. To ensure convergence, we tested several neighbor interaction cutoffs. The q-mesh density, which is the phonon momentum space sampling grid, was set to $10 \times 10 \times 10$ for most materials. We examined the convergence of grid density for all cases, and the reported values of thermal conductivity are all converged.

% For the high-throughput screening for materials with an ultralow thermal conductivity, 700 candidate materials were obtained from the Materials Project~\cite{jain2013commentary}. We limited the range of search to stable and experimentally existent binary semiconductors with a band gap of 0 to 3~eV.

\section{Results and Discussion}

\subsection{Anti-bonding valence bands in \ce{PbTe} and \ce{CsPbBr3}}

\begin{figure}[tbp] %!htb
\includegraphics[width=1\textwidth]{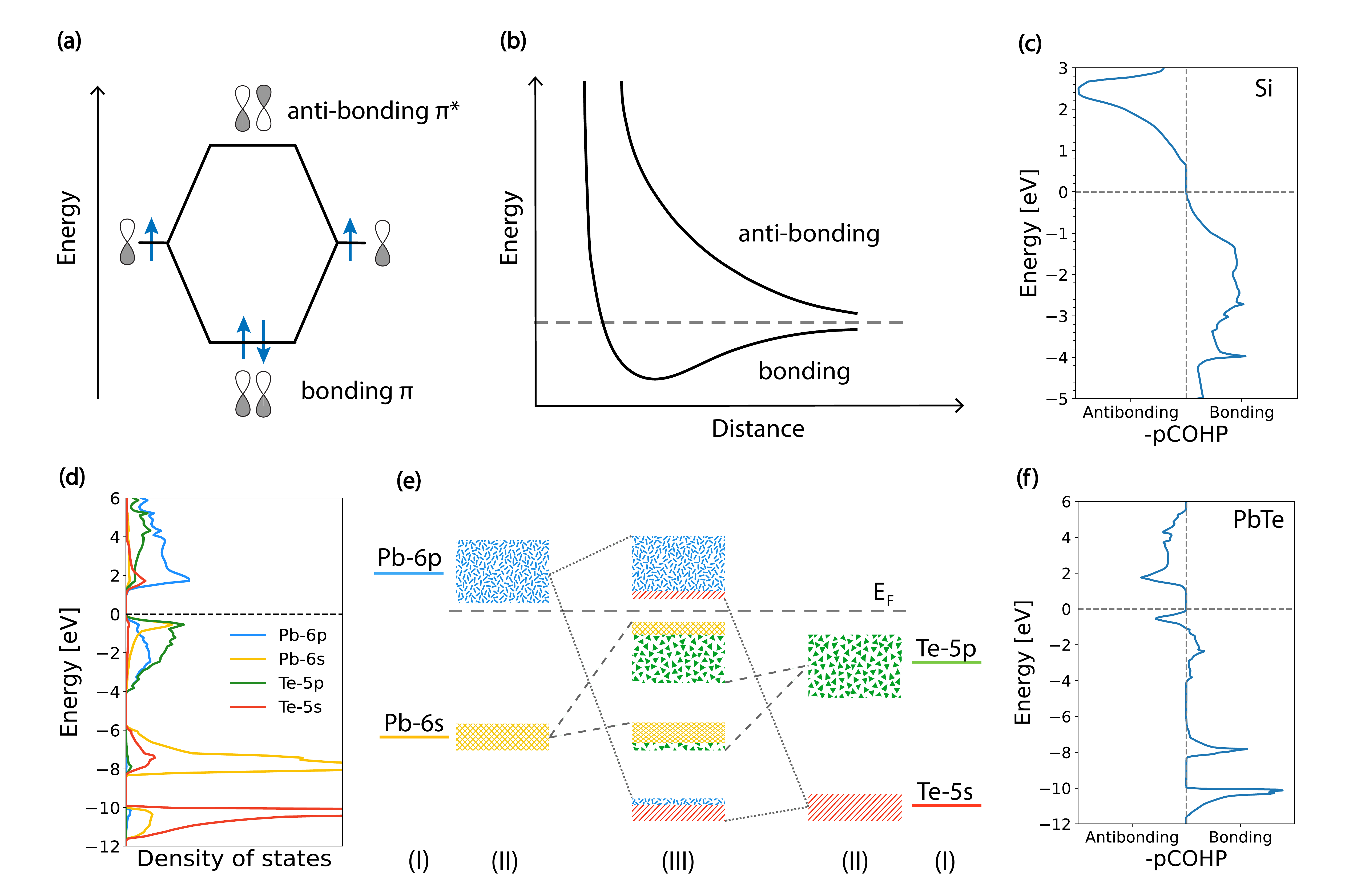}
\caption{\textbf{Schematics of anti-bonding states and anti-bonding valence bands.}
(a) Molecular orbital diagram of a $\pi$ bond formed by p orbitals.
(b) Energy of typical bonding and anti-bonding orbitals as a function of atomic distances.
(c) COHP diagram of \ce{silicon}, showing the bonding nature of the valence band and the anti-bonding nature of the conduction band.
(d) Electronic density of states of \ce{PbTe}, projected onto atomic orbitals.
(e) Molecular orbital diagram of \ce{PbTe}, showing three stages of orbital interactions: (\uppercase\expandafter{\romannumeral1}) isolated Pb and Te atomic levels; (\uppercase\expandafter{\romannumeral2}) broadened atomic levels due to s-s and p-p mixing;
(\uppercase\expandafter{\romannumeral3}) s-p hybridization introduces s orbital features to valence band maximum (VBM) and conduction band minimum (CBM) and an occupied anti-bonding valence band below the Fermi level $E_F$.
(f) COHP diagram of \ce{PbTe}, showing the strong anti-bonding nature of the valence band.
} 
\label{fig:fig1}
\end{figure}

In the chemical bonding theory~\cite{pople1970molecular,hehre1976ab}, molecular orbitals~(MOs) are formed by the overlap of atomic orbitals.
For example, Fig.~\ref{fig:fig1}(a) illustrates the $\pi$ bond formation of atomic $p$ orbitals between two adjacent atoms.
The overlap of electron wave functions splits the original atomic $p$ orbital into two MOs.
The lower energy orbital is the bonding MO, shown as the ``bonding $\pi$'' in Fig.~\ref{fig:fig1}(a). 
The higher energy orbital is the anti-bonding MO, shown as the ``anti-bonding $\pi^\ast$''.
Due to the Pauli exclusion principle, the two electrons will occupy the two spin states of the bonding MO, leaving the anti-bonding MO unoccupied. 
Consider the energy level of the two MOs as a function of the atomic distance, as shown in Fig.~\ref{fig:fig1}(b). 
 The occupation of the bonding MO lowers the total energy near the equilibrium interatomic distance and results in stabilization of the chemical bond, while the occupation of the anti-bonding MO raises the energy and, thus, weakens the bond and potentially increases the bond anharmonicity. Many of the covalently bonded semiconductors possess a valence band with occupied bonding states and a conduction band with empty anti-bonding states. For example, the valence band and the conduction band in silicon are formed by the bonding and anti-bonding states of the $sp^3$ hybridized orbitals, respectively. This can be clearly seen from the COHP analysis of silicon, which is shown in Fig.~\ref{fig:fig1}(c). 

 In contrast, the bonding nature of the electronic bands in \ce{PbTe} is qualitatively different. PbTe has been extensively studied due to its exceptional thermoelectric properties including a high Seebeck coefficient and a low thermal conductivity~\cite{snyder2008complex,heremans2008enhancement,akhmedova2009effect}.
Despite its high-symmetry structure, the bulk lattice thermal conductivity in single-crystalline PbTe is unexpectedly low, around 2\,W/(m K) at room temperature~\cite{akhmedova2009effect}.
Fig.~\ref{fig:fig1}(d) shows the PbTe electronic density of states (DOS) that has been projected onto Pb-6p, Pb-6s, Te-5p, and Te-5s orbitals. Based on the electronic DOS, Fig.~\ref{fig:fig1}(e) shows a simplified linear combination of atomic orbitals (LCAO) picture of PbTe. Pb$^{2+}$ ions have empty 6p states, but the 6s states are occupied by a ``lone pair'' of electrons~\cite{waghmare2003first}. This unique ``lone pair'' configuration enables efficient s-p mixing between the Pb-6s states and the Te-5p states, forming bonding and anti-bonding states. Since both Pb-6s and Te-6p states are fully occupied, both the resulted bonding and anti-bonding states are occupied, leading to the highest-occupied valence band possessing a strong anti-bonding character, as shown in the COHP analysis shown in Fig.~\ref{fig:fig1}(f). This intimate relationship between the lone-pair electrons of the divalent group-IV cations and the ABVB was also discussed in previous studies~\cite{waghmare2003first}. However, a direct evaluation of the impact of the ABVB on the thermal transport is still lacking.

To quantify the ABVB effect on thermal transport properties in PbTe, we hypothesized that, by removing electrons from the anti-bonding states at the VBM and then relaxing the lattice, the chemical bond strength will be increased and the bond anharmonicity will be reduced, leading to a stabilized lattice and an increased thermal conductivity. To test this hypothesis, we used a $4\times4\times4$ supercell in the calculation, where electrons were removed in \ce{PbTe} and the lattice was fully relaxed afterwards.
We performed the calculation by removing 2 and 8 electrons (labeled ``-2e'' and ``-8e'' cases, respectively, in the following discussion) out of a total number of 640 valence electrons, which can be considered a small perturbation to the pristine PbTe electronic structure and only represent the anti-bonding states very close to the VBM.
A 4-fold degeneracy exists for 8 VBM electrons located at L (0.5, 0.5, 0.5) points, which are shifted to the $\Gamma$ point in the supercell calculation.
Therefore, a single $\Gamma$ point sampling was sufficient in the $4\times4\times4$ supercell.

% \ce{PbTe} has 8 electrons at VBM with 4-fold degeneracy located at L (0.5, 0.5, 0.5) point.
% We consider three cases when removing electrons: 
% $\langle 1 \rangle$ -2e unrelaxed, remove two electrons and bond length is fixed without lattice relaxation, 
% $\langle 2 \rangle $ -2e relaxed, remove two electrons and bond length is allowed to change with lattice relaxation. 
% $\langle 3 \rangle $ -8e relaxed, remove eight electrons and allow lattice relaxation. 
% The implementation details can be found in Sec.~\ref{sec:method}.

\begin{figure}[tbp] %!htb
\includegraphics[width=1\textwidth]{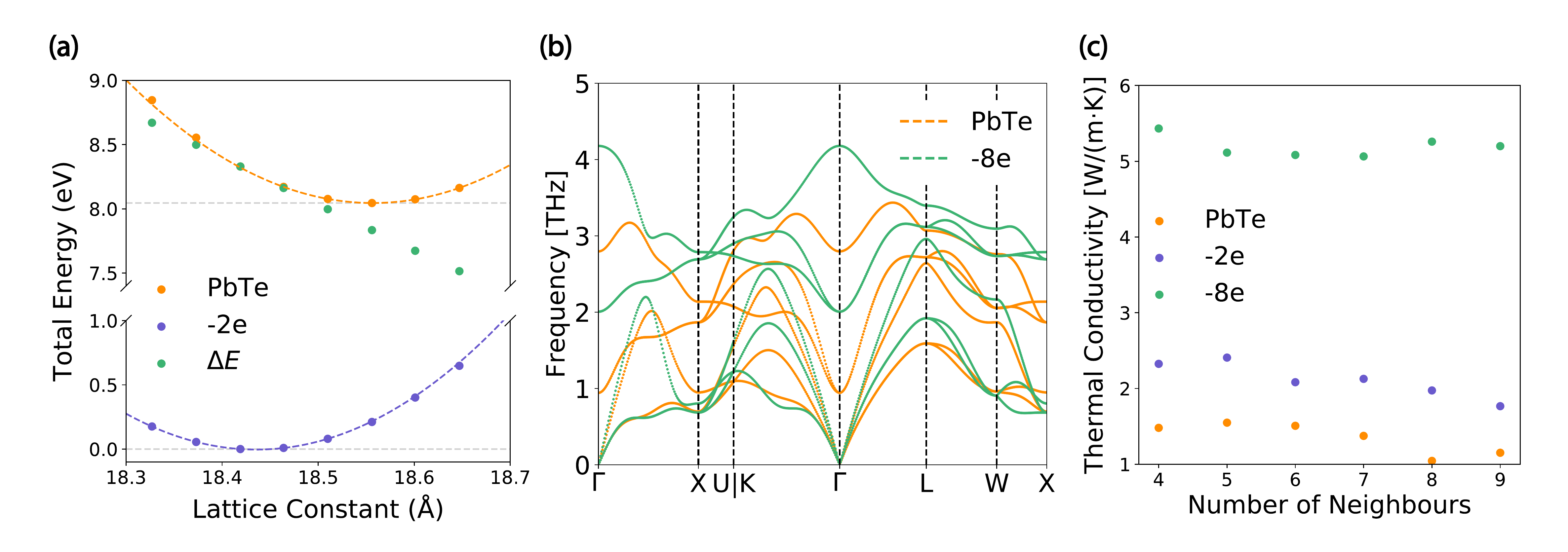}
\caption{\textbf{Impact of the anti-bonding valence band on thermal transport in PbTe.}
(a) Total ground-state energy of unperturbed \ce{PbTe} and the ``-2e'' case as a function of the supercell lattice constant. The difference of the two energies, $\Delta E$ is also shown, which reflects the contribution of the removed anti-bonding states near the VBM. The energy axis is shifted such that the energy of the ``-2e'' case at the equilibrium volume is zero.
(b) Calculated phonon dispersion relations of the unperturbed PbTe and the ``-8e'' case, showing the impact of anti-bonding states on the acoustic phonon velocity and the soft optical phonon frequency.(c) Calculated lattice thermal conductivity of the unperturbed PbTe and the ``-2e'' and ``-8e'' cases as a function of the interacting neighbor shells included in the calculation, showing the significant impact of the occupied anti-bonding valence states on the thermal conductivity of PbTe.
} 
\label{fig:fig2}
\end{figure}

%, corresponding to the removal of anti-bonding energy curve in Fig.~\ref{fig:01_anti-bonding}(b)
% The simplest bonding picture is to only consider the nearest neighbour bonds. The Pb and Te atoms are connected by springs, and the bonding strength can be represented by the ``spring constants'' (force constants).
% The spring constant for a chemical bond is the parabola curvature in energy curves shown in Fig.~\ref{fig:fig1}(b), and can be calculated by the second order coefficients in Taylor expansion.
% Larger force constants mean stiffer bonds and stronger interactions.
The calculated IFCs and the bond lengths are shown in Table~\ref{tab:ifc}, where we compare three cases: unperturbed PbTe, the ``-2e'' case and the ``-8e'' case. Here, $K_{12}$ is the trace of the IFC tensor of the nearest Pb-Te bond. The negative sign of the $K_{12}$ values indicates the overall stable bonding character of the Pb-Te bond and their magnitude reflects the bonding strength. From Table~\ref{tab:ifc}, it is clearly seen that removing electrons from the ABVB states decreases the length of the Pb-Te bond and increases its bonding strength, thus stabilizing the PbTe lattice. 
This effect is further illustrated in Fig.~\ref{fig:fig2}(a) as the total ground-state energy of the unperturbed PbTe and the ``-2e'' case is plotted as a function of the lattice constant. The minimum location, corresponding to the equilibrium lattice parameter, shifts towards a smaller value in the ``-2e'' case, with an increased quadratic coefficient from 0.0304 to 0.0319, indicating a strengthened bond. The difference in the ground-state energy between the two cases, $\Delta E$, reflects the contribution from the anti-bonding component, showing a destabilizing trend.

\begin{table}[htb]
\caption{\label{tab:ifc}
Traces of the IFC Tensor and Bond Length of \ce{PbTe}
}
\begin{ruledtabular}
\begin{tabular}{c|cccc}
IFC & PbTe &  -2e &  -8e \\
\hline
\hline
K$_{12}$ & -0.786 & -0.869 & -1.18
\\
\hline
Bond Length & 3.28\AA & 3.26\AA & 3.19\AA
\\
% 3.280, 3.257, 3.191
\end{tabular}
\end{ruledtabular}
\end{table}

The impact of removing electrons from ABVB states on phonon properties is examined in Fig.~\ref{fig:fig2}(b), which illustrates the phonon dispersions of the unperturbed \ce{PbTe} and the ``-8e'' case.
Firstly, the acoustic phonon modes have an increased velocity in the ``-8e'' case compared to that in the unperturbed case, which is a consequence of the increased bonding strength consistent with the magnitude of $K_{12}$ in Table~\ref{tab:ifc}. Secondly, removing electrons from the ABVB states significantly raises the frequency of the soft optical phonons near the $\Gamma$ point, suggesting that the occupied ABVB in \ce{PbTe} not only weakens the chemical bond, but also leads to long-ranged force interactions resulting in soft optical phonons. Both effects are expected to strongly affect the thermal conductivity.
Fig.~\ref{fig:fig2}(c) shows the calculated thermal conductivity in unperturbed PbTe as compared to ``-2e'' and ``-8e'' cases as a function of the number of neighbor shells included in calculating the third-order IFCs. 
The corresponding scattering properties of PbTe are included in the Supplementary Material~\cite{SM}.
The thermal conductivity in the ``-2e'' case is approximately 1.5 times that of the unperturbed PbTe, while in the ``-8e'' case, the thermal conductivity is boosted by at least a factor of 3. 
These findings indicate that the presence of occupied ABVBs is the origin for the abnormally low lattice thermal conductivity in PbTe from a chemical-bonding-theory point of view.
To solidify these findings, we further applied the same strategy (removing valence electrons) to Si with bonding valence bands, and the results can be found in the Supplementary Material~\cite{SM}. 
In contrast to PbTe with ABVB, removing valence electrons in the bonding states in Si leads to a weakened bond strength, a decrease in the speed of sound and a reduction in the thermal conductivity.
This result further establishes the impact of ABVB on the intrinsic lattice thermal conductivity.

\begin{figure}[tbp] %!htb
\includegraphics[width=1\textwidth]{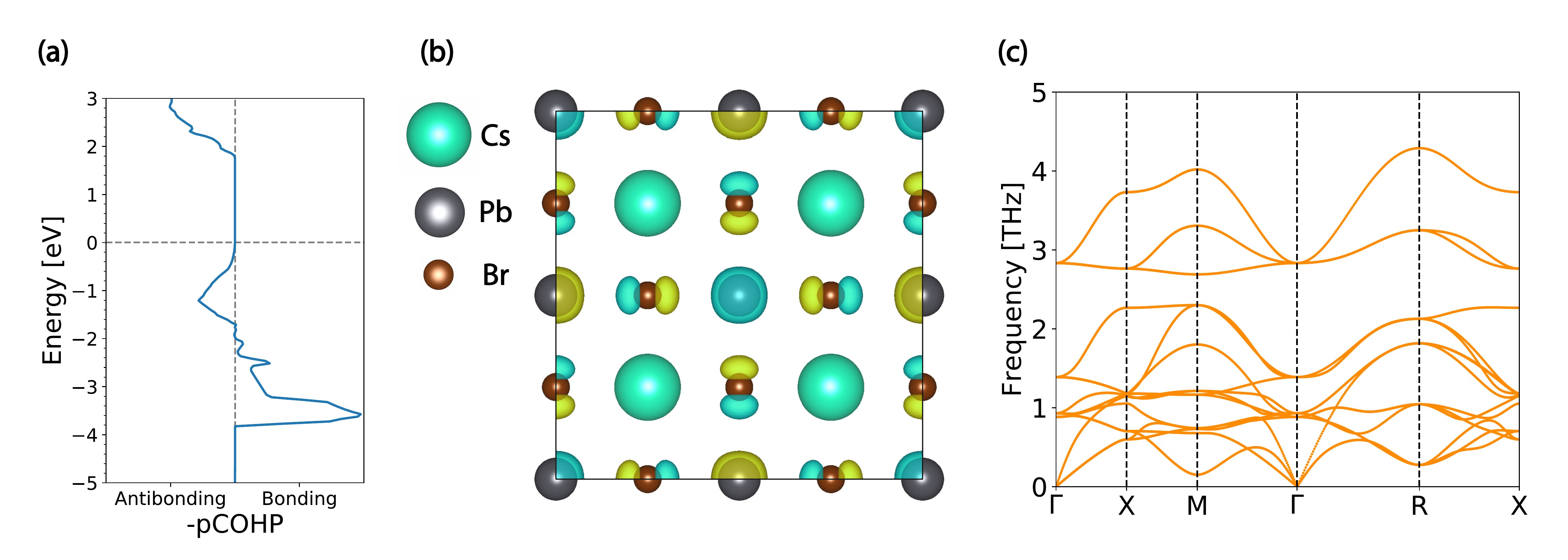}
\caption{\textbf{Anti-bonding valence bands in \ce{CsPbBr3} and their impact on phonons.}
(a) The COHP diagram for \ce{CsPbBr3}, where a strong anti-bonding feature is shown for the VBM.
(b) Isosurfaces of the electronics wavefunctions at the VBM in \ce{CsPbBr3}. The s orbitals on Pb atoms and p orbitals on Br atoms show anti-bonding interactions as reflected by the opposite signs of the wavefunctions associated with facing Pb and Br atomic pairs.
(c) The calculated phonon dispersion of the cubic phase of \ce{CsPbBr3} at 400\,K. Low acoustic velocity associated with the anti-bonding valence bands is observed, which drives the low intrinsic thermal conductivity.
} 
\label{fig:fig3}
\end{figure}

Similarly, we also examined the relationship between the ABVB and the thermal transport properties of the inorganic halide perovskite \ce{CsPbBr3} that also contains the Pb$^{2+}$ ion.
Halide perovskites have gained attention as a new class of photovoltaic and thermoelectric material \cite{jung2019efficient,li2022organometallic,kim2020high,haque2020halide}. 
While the ultralow thermal conductivity of organic-inorganic hybrid halide perovskites was often attributed to the cation dynamic disorder, it is not well understood why thermal transport in crystalline all-inorganic halide perovskites such as \ce{CsPbBr3} is also significantly suppressed~\cite{wang2018cation,xie2020all}. Similar to PbTe, the Pb$^{2+}$ ion in \ce{CsPbBr3} hosts Pb-6s lone pair electrons~\cite{seshadri1999visualizing,raulot2002ab,stoltzfus2007structure} that promote the s-p mixing between Pb-6s and Br-4p orbitals, leading to an ABVB in \ce{CsPbBr3} that can be quantified by the COHP analysis shown in Fig.~\ref{fig:fig3}(a). A visualization of the Pb-Br anti-bonding interaction is shown in Fig.~\ref{fig:fig3}(b), where isosurfaces of the electronic wavefunctions at the VBM of \ce{CsPbBr3} are shown. Here, blue and yellow isosurfaces indicate opposite signs of the wavefunction and the opposite signs of the isosurfaces associated with facing Pb and Br pairs confirm the anti-bonding nature of the valence band in \ce{CsPbBr3}. Figure~\ref{fig:fig3}(c) shows the phonon dispersion relation of the cubic phase \ce{CsPbBr3} at 400\,K simulated by the temperature-dependent effective potential (TDEP) method~\cite{hellman2013temperature}, where the low acoustic phonon velocity originated from the weakened bonds due to the ABVBs is clearly seen that drive the low lattice thermal conductivity. 

\subsection{High-throughput Screening of Semiconductors with ABVBs}\label{subsec:all_material}

Results from \ce{PbTe} and \ce{CsPbBr3} suggest that a strong anti-bonding character in the highest occupied valence band, which can be efficiently evaluated by the COHP analysis, is a convenient indicator for weakened covalent bonds, large phonon anharmonicity and a low lattice thermal conductivity.
This motivates us to perform a high-throughput screening for semiconductors with strong ABVBs in the Materials Project database~\cite{jain2013commentary} by the COHP analysis. As a first step, we focused on binary compounds and found $\sim$1000 binary candidates in the Materials Project database that are stable and have a finite band gap between 0 to 3 eV. Materials containing heavy elements with unfilled $f$ electrons were excluded from the candidate list due to the known inaccuracy of DFT to deal with these materials.
For each candidate, we calculated COHP for all pair-wise interactions within 1.5 times the nearest-neighbor distance.
Then the strength of the anti-bonding character was quantified by integrating the area under the COHP curves within 0.1\,eV of the valence band edge, thus only focusing on the highest-occupied valence band states.
All candidates were sorted by the strength of the anti-bonding character associated with their highest occupied valence band states.
A screened list consisting of 625 experimentally stable binary semiconductors are provided in Table~S2 of the Supplementary Material~\cite{SM}.
% From our candidate list, we identified that strong anti-bonding interactions often exist between reactive non-metals, e.g. O-O in \ce{SrO2}, S-S in \ce{KS} and Cl-Cl in \ce{NaCl3}, which have ``over-filled covalent bonds'' so that the higher energy anti-bonding orbitals are also occupied.
Among these materials of interest, we present a comprehensive study of the thermal transport of the \ce{XS} family, where X = Na, K, Rb, Cs are alkaline metals.
Their structural properties and thermal conductivity at room temperature are summarized in Table~\ref{tab:xs}, and the convergence tests of their thermal conductivities are provided in the Supplementary Material~\cite{SM}. As shown in Table~\ref{tab:xs}, with an increasing mass of the alkaline metal, the room-temperature thermal conductivity decreases from 5\,W/(m K) in \ce{NaS} to 0.1\,W/(m K) in \ce{CsS}.
\ce{KS}, \ce{RbS}, \ce{CsS} all show an ultralow room-temperature thermal conductivity $<$ 0.8\,W/(m K) despite their relatively simple crystal structures.
The results open up possibilities for their potential applications in thermoelectrics and also motivate further experimental studies.
Notably, \ce{KS} reaches an ultralow thermal conductivity of 0.8\,W/(m K) at room temperature without heavy elements or complex structures.
A detailed analysis of the bonding mechanism and the thermal transport was thus carried out for \ce{KS} as an example to demonstrate the decisive impact of the ABVB on its ultralow thermal conductivity.

\begin{table}[htb]
\caption{\label{tab:xs}
Basic properties and calculated thermal conductivity of NaS, KS, RbS, CsS
}
\begin{ruledtabular}
\begin{tabular}{c|cccccccc}
Material & Band Gap [eV] & Crystal System & Space Group & ${\kappa}_{x}$ & ${\kappa}_{y}$ & ${\kappa}_{z}$ & [W/(m K)]
\\
\hline
\hline
NaS & 1.23 & hexagonal & P$6_{3}$/mmc & 5.9 & 5.9 & 3.9  % 8
\\
\hline
KS & 1.47 & hexagonal & P$\bar{6}2$m & 0.84 & 0.84 & 0.76% 12
\\
\hline
RbS & 1.58 & hexagonal & P$\bar{6}2$m &0.65 & 0.65 & 0.64 % 12 
\\
\hline
CsS & 1.73 & orthorhombic & Immm & 0.15 & 0.14 & 0.083 % 4
\\
\end{tabular}
\end{ruledtabular}
\end{table}

\subsection{Detailed Analysis of Bonding and Thermal Transport in KS}\label{subsec:KS}

\begin{figure}[tbp] %!htb
\includegraphics[width=1\textwidth]{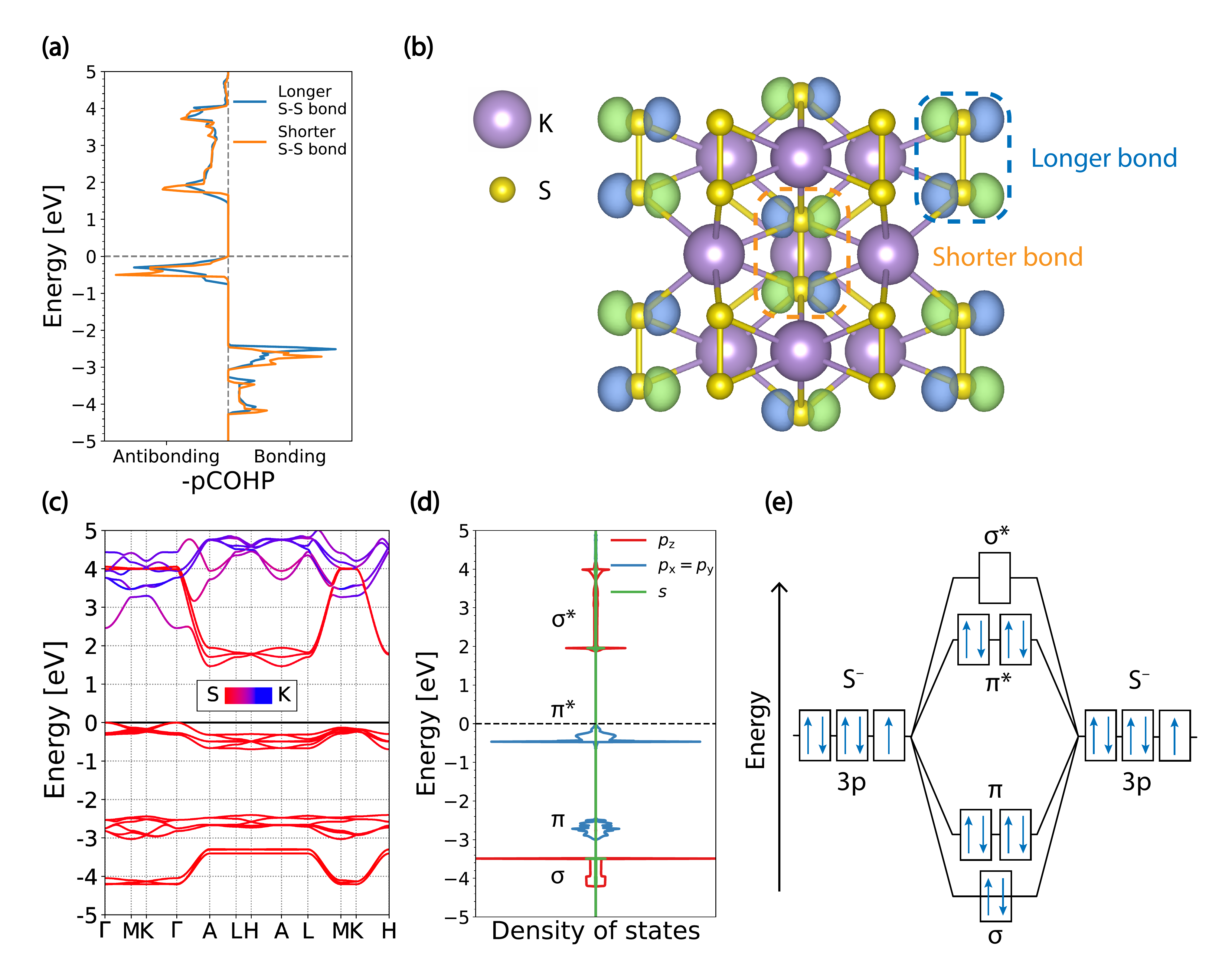}
\caption{\textbf{Anti-bonding valence bands in \ce{KS}.}
(a) The COHP diagram of \ce{KS}, where the S-S bonds show strong anti-bonding features near the VBM. The longer and shorter S-S bonds correspond to a bond length of 2.15~\AA~and 2.13~\AA, respectively.
(b) The isosurfaces of the VBM electronic wavefunctions in \ce{KS}. The longer and shorter S-S bonds are marked in blue and orange boxes, respectively.
(c) The calculated electronic band structure of \ce{KS}. The S orbitals contribute dominantly to the valence band.
(d) $l$-decomposed and site-projected electronic DOS for S orbitals in \ce{KS}. $p_x$ (equivalent to $p_y$) orbitals form the valence bands, and the highest occupied valence band consists of the anti-bonding $\pi^\ast$ state formed by $p_x$ ($p_y$) orbitals.
(e) Molecular bond analysis for the S-S bond.
The 3p orbitals of a single S$^{-}$ ion has 5 electrons. The covalent S-S bond form $\sigma, \sigma^\ast, \pi$ and $\pi^\ast$ orbitals, where $\sigma, \pi$ and $\pi^\ast$ bonds are occupied.
These occupied bonds correspond to the electronic DOS below the Fermi level as shown in Fig.~\ref{fig:fig4}(d) and the highest occupied valence band is composed of $\pi^\ast$ anti-bonding states.
} 
\label{fig:fig4}
\end{figure}

\begin{figure}[tbp] %!htb
\includegraphics[width=1\textwidth]{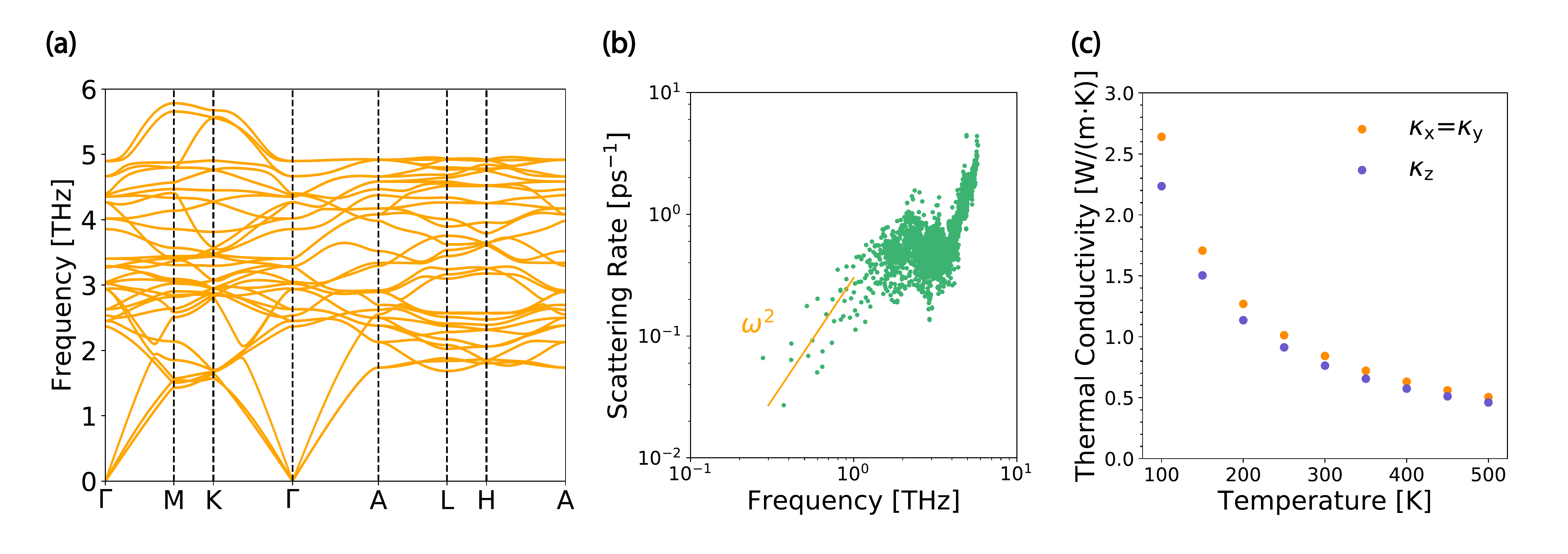}
\caption{\textbf{Calculated thermal transport properties of \ce{KS}.}
(a) The calculated phonon dispersion relation of \ce{KS}, which shows the material is dynamically stable. Flat phonon bands are also observed near 1.5\,THz.
(b) The calculated phonon-phonon scattering rate in \ce{KS} at 300\,K. The scattering rates of the low-frequency acoustic phonons follow the frequency-square scaling, while flat phonon bands lead to increased scattering near 1.5\,THz.
(c) The calculated temperature-dependent lattice thermal conductivity of \ce{KS} from 100\,K to 500\,K, showing an ultralow lattice thermal conductivity below 1\,W/(m K) at room temperature.
} 
\label{fig:fig5}
\end{figure}

Despite its unusual chemical formula, \ce{KS} (or \ce{K2S2}) has been theoretically found to be as stable as \ce{K2S} at room temperature under ambient pressure~\cite{li2017unexpected} and has been experimentally synthesized~\cite{sangster1997ks,bottcher1993kenntnis}.
Building upon these results, we conducted a more in-depth analysis of \ce{KS}, focusing specifically on its chemical bonding properties.
The COHP diagram shown in Fig.~\ref{fig:fig4}(a) attributes the strong anti-bonding character of the valence band to two kinds of S-S bonds. These two S-S bonds can be visualized in the isosurfaces of VBM electron wavefunctions as shown in Fig.~\ref{fig:fig4}(b), where the blue and yellow isosurfaces represent opposite signs of the wavefunction. From the isosurfaces, the two anti-bonding S-S bonds are $\pi^\ast$ bonds of the S-3p orbitals, where the neighboring wavefunctions possess opposite signs.
The longer bond (2.15~\AA), marked in a blue box in Fig.~\ref{fig:fig4}(b), has more prominent anti-bonding component closer to the VBM. 
The shorter bond labeled in an orange box shows stronger anti-bonding due to its shorter (2.13~\AA) length.
% The S atoms have $p$ orbitals, and the $p$ orbitals are antisymmetrized in the S-S bonds, indicating the anti-bonding $\pi^\ast$ bond in Fig.~\ref{fig:fig1}(a). 
The calculated electronic band structure in Fig.~\ref{fig:fig4}(c) shows that orbitals from the S atom make dominant contributions to electronic states near the band edges, while K orbitals only contribute to unoccupied higher energy conduction bands. 
The effective masses of electrons and holes near the band edges can also be extracted from the calculated band structure: the electron effective mass is $0.24 m_0$ along the out-of-plane direction and $1.48 m_0$ along the in-plane direction; the hole effective mass is $1.2 m_0$ along the out-of-plane direction and $3.4 m_0$ along the in-plane direction, where $m_0$ is the free electron mass. The relatively low effective masses show that the electronic transport properties are maintained despite the reduced bonding strength leading to a lower thermal conductivity.    
% add more comments about bs
%shows that VBM is at $\Gamma$ point and CBM is at $A$ point.
From the orbital-decomposed and site-projected electronic DOS for an S atom shown in Fig.~\ref{fig:fig4}(d), we can observe that the equivalent $p_x$ and $p_y$ orbitals form the anti-bonding $\pi^\ast$ state that corresponds to the highest occupied valence band in KS.
The molecular orbital diagram of the S-S bond in KS is illustrated in Fig.~\ref{fig:fig4}(e) to reveal the origin of the ABVB in KS. 
The outermost 4s electron in a K atom is transferred to an S atom to form a $S^{-}$ ion in its unusual monovalent configuration.
The 3p orbitals of a single $S^{-}$ ion thus contain 5 electrons. 
The covalent S-S bond form $\sigma, \sigma^\ast, \pi$ and $\pi^\ast$ orbitals, where $\sigma, \pi$ and $\pi^\ast$ orbitals are occupied, yielding a bond order of 1. Here the anti-bonding $\pi^\ast$ state is occupied thanks to the additional electron transferred from the K atom.
These occupied orbitals correspond to the electronic DOS below the Fermi level in Fig.~\ref{fig:fig4}(e).
Therefore, the highest occupied valence band is composed of anti-bonding $\pi^\ast$ states. This mechanisms is in addition to the group-IV lone-pair electrons that can give rise to ABVBs. Recently, He et al. also pointed out that p-d hybridization can lead to ABVBs and ultralow thermal conductivities in Cu- and Ag-containing compounds~\cite{he2022accelerated}. These chemical insights into ABVB formation can further guide the search and design of new materials with an ultralow intrinsic thermal conductivity.
% By this argument, to search for strong valence band anti-bonding, the candidate should have covalent bond such that the highest energy $\sigma^\ast$ bond is occupied.
% \ce{NaCl3} is a good example of such a candidate and ranks high on our list. 
% The $[Cl-Cl-Cl]^{-}$ cluster in \ce{NaCl3} has two covalent bonds, and the bond order of the Cl-Cl bond is less than 1.

To complete our discussion, we report the calculated thermal transport properties of \ce{KS} in Fig.~\ref{fig:fig5}. Figure~\ref{fig:fig5}(a) shows the phonon dispersion, suggesting that \ce{KS} is dynamically stable. More importantly, flat phonon bands can be observed near 1.5 THz, which are expected to scatter acoustic phonons and hinder thermal transport. Figure~\ref{fig:fig5}(b) depicts the phonon-phonon scattering rates as a function of the phonon frequency, which are compared to the expected frequency-square scaling. The scattering rates of the low-frequency acoustic phonons follow the frequency-square trend while the flat phonon bands lead to a peak in the scattering rates near 1.5 THz . The calculated lattice thermal conductivity of \ce{KS} is 0.8\,W/(m K) at room temperature, which is lower than most compounds with similar atomic mass and simple crystals structure. The temperature-dependent thermal conductivity of \ce{KS} from 100\,K to 500\,K is displayed in Figure~\ref{fig:fig5}(c). 
The compound \ce{CsS} in the same family with a heavier alkaline metal even possesses a room-temperature thermal conductivity as low as 0.1\,W/(m K) (more details are provided in the Supplementary Material~\cite{SM}), even lower than many amorphous and organic materials. These materials are promising candidates for applications where an ultralow thermal conductivity is desired.

\section{Conclusion}\label{sec:conc}
In conclusion,
we examined the connection between ABVBs and low thermal conductivities with first-principles calculations in \ce{PbTe} and \ce{CsPbBr3} and we found that the ABVBs are responsible for the weakened chemical bonds and the soft optical phonons that lead to abnormally low lattice thermal conductivities in both compounds. Based on this observation, we conducted a high throughput materials search based on ABVBs and found over 600 experimentally stable binary semiconductors with strong ABVBs.
Among the candidate materials, we analyzed in detail the XS (X = Na, K, Rb, and Cs) family in terms of their strong ABVB states and evaluated their lattice thermal conductivities. 
With an ultralow thermal conductivity of 0.1\,W/(m K), \ce{CsS} is an exceptional example of a crystalline material with a lower thermal conductivity than most amorphous and organic materials. Several other materials on our list with strong ABVBs are potentially interesting as well, such as \ce{NaCl3}, and further studies are needed to fully evaluate the impact of the ABVBs on their electrical and thermoelectric transport properties. Our material search can also be expanded to more complicated materials containing three or more elements. Our results suggest that the intuitions offered by the chemical bonding theory can provide simple but powerful guidelines for understanding the thermal conductivity of materials as well as for discovering new materials with unusual thermal transport properties.

\begin{acknowledgments}
We acknowledge Fanghao Zhang for fruitful discussions. This work is based on research supported by the U.S. Office of Naval Research under the award number N00014-22-1-2262. Y.C. also acknowledges the support from the Graduate Traineeship Program of the NSF Quantum Foundry via the Q-AMASE-i program under award number DMR-1906325 at the University of California, Santa Barbara (UCSB). This work used Stampede2 at Texas Advanced Computing Center (TACC) and Expanse at San Diego Supercomputer Center (SDSC) through allocation MAT200011 from the Advanced Cyberinfrastructure Coordination Ecosystem: Services \& Support (ACCESS) program, which is supported by National Science Foundation grants 2138259, 2138286, 2138307, 2137603, and 2138296. 
Use is also made of computational facilities purchased with funds from the National Science Foundation (award number CNS-1725797) and administered by the Center for Scientific Computing (CSC) at University of California, Santa Barbara (UCSB). 
The CSC is supported by the California NanoSystems Institute and the Materials Research Science and Engineering Center (MRSEC; NSF DMR-1720256) at UCSB. 
\end{acknowledgments}

%
% ****** End of file apssamp.tex ******

%\bibliography{references.bib}% Produces the bibliography via BibTeX.

\begin{thebibliography}{81}%
\makeatletter
\providecommand \@ifxundefined [1]{%
 \@ifx{#1\undefined}
}%
\providecommand \@ifnum [1]{%
 \ifnum #1\expandafter \@firstoftwo
 \else \expandafter \@secondoftwo
 \fi
}%
\providecommand \@ifx [1]{%
 \ifx #1\expandafter \@firstoftwo
 \else \expandafter \@secondoftwo
 \fi
}%
\providecommand \natexlab [1]{#1}%
\providecommand \enquote  [1]{``#1''}%
\providecommand \bibnamefont  [1]{#1}%
\providecommand \bibfnamefont [1]{#1}%
\providecommand \citenamefont [1]{#1}%
\providecommand \href@noop [0]{\@secondoftwo}%
\providecommand \href [0]{\begingroup \@sanitize@url \@href}%
\providecommand \@href[1]{\@@startlink{#1}\@@href}%
\providecommand \@@href[1]{\endgroup#1\@@endlink}%
\providecommand \@sanitize@url [0]{\catcode `\\12\catcode `\$12\catcode
  `\&12\catcode `\#12\catcode `\^12\catcode `\_12\catcode `\%12\relax}%
\providecommand \@@startlink[1]{}%
\providecommand \@@endlink[0]{}%
\providecommand \url  [0]{\begingroup\@sanitize@url \@url }%
\providecommand \@url [1]{\endgroup\@href {#1}{\urlprefix }}%
\providecommand \urlprefix  [0]{URL }%
\providecommand \Eprint [0]{\href }%
\providecommand \doibase [0]{https://doi.org/}%
\providecommand \selectlanguage [0]{\@gobble}%
\providecommand \bibinfo  [0]{\@secondoftwo}%
\providecommand \bibfield  [0]{\@secondoftwo}%
\providecommand \translation [1]{[#1]}%
\providecommand \BibitemOpen [0]{}%
\providecommand \bibitemStop [0]{}%
\providecommand \bibitemNoStop [0]{.\EOS\space}%
\providecommand \EOS [0]{\spacefactor3000\relax}%
\providecommand \BibitemShut  [1]{\csname bibitem#1\endcsname}%
\let\auto@bib@innerbib\@empty
%</preamble>
\bibitem [{\citenamefont {Mao}\ \emph {et~al.}(2018)\citenamefont {Mao},
  \citenamefont {Liu}, \citenamefont {Zhou}, \citenamefont {Zhu}, \citenamefont
  {Zhang}, \citenamefont {Chen},\ and\ \citenamefont {Ren}}]{mao2018advances}%
  \BibitemOpen
  \bibfield  {author} {\bibinfo {author} {\bibfnamefont {J.}~\bibnamefont
  {Mao}}, \bibinfo {author} {\bibfnamefont {Z.}~\bibnamefont {Liu}}, \bibinfo
  {author} {\bibfnamefont {J.}~\bibnamefont {Zhou}}, \bibinfo {author}
  {\bibfnamefont {H.}~\bibnamefont {Zhu}}, \bibinfo {author} {\bibfnamefont
  {Q.}~\bibnamefont {Zhang}}, \bibinfo {author} {\bibfnamefont
  {G.}~\bibnamefont {Chen}},\ and\ \bibinfo {author} {\bibfnamefont
  {Z.}~\bibnamefont {Ren}},\ }\bibfield  {title} {\bibinfo {title} {Advances in
  thermoelectrics},\ }\href {https://doi.org/10.1080/00018732.2018.1551715}
  {\bibfield  {journal} {\bibinfo  {journal} {Adv. Phys.}\ }\textbf {\bibinfo
  {volume} {67}},\ \bibinfo {pages} {69} (\bibinfo {year} {2018})}\BibitemShut
  {NoStop}%
\bibitem [{\citenamefont {Chen}\ \emph {et~al.}(2018)\citenamefont {Chen},
  \citenamefont {Zhang},\ and\ \citenamefont {Pei}}]{chen2018manipulation}%
  \BibitemOpen
  \bibfield  {author} {\bibinfo {author} {\bibfnamefont {Z.}~\bibnamefont
  {Chen}}, \bibinfo {author} {\bibfnamefont {X.}~\bibnamefont {Zhang}},\ and\
  \bibinfo {author} {\bibfnamefont {Y.}~\bibnamefont {Pei}},\ }\bibfield
  {title} {\bibinfo {title} {Manipulation of phonon transport in
  thermoelectrics},\ }\href {https://doi.org/10.1002/adma.201705617} {\bibfield
   {journal} {\bibinfo  {journal} {Adv. Mater.}\ }\textbf {\bibinfo {volume}
  {30}},\ \bibinfo {pages} {1705617} (\bibinfo {year} {2018})}\BibitemShut
  {NoStop}%
\bibitem [{\citenamefont {Qian}\ \emph {et~al.}(2021)\citenamefont {Qian},
  \citenamefont {Zhou},\ and\ \citenamefont {Chen}}]{qian2021phonon}%
  \BibitemOpen
  \bibfield  {author} {\bibinfo {author} {\bibfnamefont {X.}~\bibnamefont
  {Qian}}, \bibinfo {author} {\bibfnamefont {J.}~\bibnamefont {Zhou}},\ and\
  \bibinfo {author} {\bibfnamefont {G.}~\bibnamefont {Chen}},\ }\bibfield
  {title} {\bibinfo {title} {Phonon-engineered extreme thermal conductivity
  materials},\ }\href {https://doi.org/10.1038/s41563-021-00918-3} {\bibfield
  {journal} {\bibinfo  {journal} {Nat. Mater.}\ }\textbf {\bibinfo {volume}
  {20}},\ \bibinfo {pages} {1188} (\bibinfo {year} {2021})}\BibitemShut
  {NoStop}%
\bibitem [{\citenamefont {Liao}\ and\ \citenamefont
  {Chen}(2015)}]{liao2015nanocomposites}%
  \BibitemOpen
  \bibfield  {author} {\bibinfo {author} {\bibfnamefont {B.}~\bibnamefont
  {Liao}}\ and\ \bibinfo {author} {\bibfnamefont {G.}~\bibnamefont {Chen}},\
  }\bibfield  {title} {\bibinfo {title} {Nanocomposites for thermoelectrics and
  thermal engineering},\ }\href {https://doi.org/10.1557/mrs.2015.197}
  {\bibfield  {journal} {\bibinfo  {journal} {MRS Bull.}\ }\textbf {\bibinfo
  {volume} {40}},\ \bibinfo {pages} {746} (\bibinfo {year} {2015})}\BibitemShut
  {NoStop}%
\bibitem [{\citenamefont {Clarke}\ and\ \citenamefont
  {Phillpot}(2005)}]{clarke2005thermal}%
  \BibitemOpen
  \bibfield  {author} {\bibinfo {author} {\bibfnamefont {D.~R.}\ \bibnamefont
  {Clarke}}\ and\ \bibinfo {author} {\bibfnamefont {S.~R.}\ \bibnamefont
  {Phillpot}},\ }\bibfield  {title} {\bibinfo {title} {Thermal barrier coating
  materials},\ }\href {https://doi.org/10.1016/s1369-7021(05)70934-2}
  {\bibfield  {journal} {\bibinfo  {journal} {Mater. Today}\ }\textbf {\bibinfo
  {volume} {8}},\ \bibinfo {pages} {22} (\bibinfo {year} {2005})}\BibitemShut
  {NoStop}%
\bibitem [{\citenamefont {Nolas}\ \emph
  {et~al.}(2001{\natexlab{a}})\citenamefont {Nolas}, \citenamefont {Slack},\
  and\ \citenamefont {Schujman}}]{nolas2001semiconductor}%
  \BibitemOpen
  \bibfield  {author} {\bibinfo {author} {\bibfnamefont {G.~S.}\ \bibnamefont
  {Nolas}}, \bibinfo {author} {\bibfnamefont {G.~A.}\ \bibnamefont {Slack}},\
  and\ \bibinfo {author} {\bibfnamefont {S.~B.}\ \bibnamefont {Schujman}},\
  }\bibfield  {title} {\bibinfo {title} {Chapter 6 semiconductor clathrates:
  {A} phonon glass electron crystal material with potential for thermoelectric
  applications},\ }in\ \href {https://doi.org/10.1016/s0080-8784(01)80152-6}
  {\emph {\bibinfo {booktitle} {Recent Trends in Thermoelectric Materials
  Research I}}},\ Vol.~\bibinfo {volume} {69}\ (\bibinfo  {publisher}
  {Elsevier},\ \bibinfo {year} {2001})\ pp.\ \bibinfo {pages}
  {255--300}\BibitemShut {NoStop}%
\bibitem [{\citenamefont {Beekman}\ \emph {et~al.}(2015)\citenamefont
  {Beekman}, \citenamefont {Morelli},\ and\ \citenamefont
  {Nolas}}]{beekman2015better}%
  \BibitemOpen
  \bibfield  {author} {\bibinfo {author} {\bibfnamefont {M.}~\bibnamefont
  {Beekman}}, \bibinfo {author} {\bibfnamefont {D.~T.}\ \bibnamefont
  {Morelli}},\ and\ \bibinfo {author} {\bibfnamefont {G.~S.}\ \bibnamefont
  {Nolas}},\ }\bibfield  {title} {\bibinfo {title} {Better thermoelectrics
  through glass-like crystals},\ }\href {https://doi.org/10.1038/nmat4461}
  {\bibfield  {journal} {\bibinfo  {journal} {Nat. Mater.}\ }\textbf {\bibinfo
  {volume} {14}},\ \bibinfo {pages} {1182} (\bibinfo {year}
  {2015})}\BibitemShut {NoStop}%
\bibitem [{\citenamefont {Takabatake}\ \emph {et~al.}(2014)\citenamefont
  {Takabatake}, \citenamefont {Suekuni}, \citenamefont {Nakayama},\ and\
  \citenamefont {Kaneshita}}]{takabatake2014phonon}%
  \BibitemOpen
  \bibfield  {author} {\bibinfo {author} {\bibfnamefont {T.}~\bibnamefont
  {Takabatake}}, \bibinfo {author} {\bibfnamefont {K.}~\bibnamefont {Suekuni}},
  \bibinfo {author} {\bibfnamefont {T.}~\bibnamefont {Nakayama}},\ and\
  \bibinfo {author} {\bibfnamefont {E.}~\bibnamefont {Kaneshita}},\ }\bibfield
  {title} {\bibinfo {title} {Phonon-glass electron-crystal thermoelectric
  clathrates: {Experiments} and theory},\ }\href
  {https://doi.org/10.1103/revmodphys.86.669} {\bibfield  {journal} {\bibinfo
  {journal} {Rev. Mod. Phys.}\ }\textbf {\bibinfo {volume} {86}},\ \bibinfo
  {pages} {669} (\bibinfo {year} {2014})}\BibitemShut {NoStop}%
\bibitem [{\citenamefont {Keppens}\ \emph {et~al.}(2000)\citenamefont
  {Keppens}, \citenamefont {Sales}, \citenamefont {Mandrus}, \citenamefont
  {Chakoumakos},\ and\ \citenamefont {Laermans}}]{keppens2000does}%
  \BibitemOpen
  \bibfield  {author} {\bibinfo {author} {\bibfnamefont {V.}~\bibnamefont
  {Keppens}}, \bibinfo {author} {\bibfnamefont {B.}~\bibnamefont {Sales}},
  \bibinfo {author} {\bibfnamefont {D.}~\bibnamefont {Mandrus}}, \bibinfo
  {author} {\bibfnamefont {B.}~\bibnamefont {Chakoumakos}},\ and\ \bibinfo
  {author} {\bibfnamefont {C.}~\bibnamefont {Laermans}},\ }\bibfield  {title}
  {\bibinfo {title} {When does a crystal conduct heat like a glass?},\ }\href
  {https://doi.org/10.1080/09500830010003830} {\bibfield  {journal} {\bibinfo
  {journal} {Phil. Mag. Lett.}\ }\textbf {\bibinfo {volume} {80}},\ \bibinfo
  {pages} {807} (\bibinfo {year} {2000})}\BibitemShut {NoStop}%
\bibitem [{\citenamefont {Snyder}\ and\ \citenamefont
  {Toberer}(2008)}]{snyder2008complex}%
  \BibitemOpen
  \bibfield  {author} {\bibinfo {author} {\bibfnamefont {G.~J.}\ \bibnamefont
  {Snyder}}\ and\ \bibinfo {author} {\bibfnamefont {E.~S.}\ \bibnamefont
  {Toberer}},\ }\bibfield  {title} {\bibinfo {title} {Complex thermoelectric
  materials},\ }\href {https://doi.org/10.1038/nmat2090} {\bibfield  {journal}
  {\bibinfo  {journal} {Nat. Mater.}\ }\textbf {\bibinfo {volume} {7}},\
  \bibinfo {pages} {105} (\bibinfo {year} {2008})}\BibitemShut {NoStop}%
\bibitem [{\citenamefont {Minnich}\ \emph {et~al.}(2009)\citenamefont
  {Minnich}, \citenamefont {Dresselhaus}, \citenamefont {Ren},\ and\
  \citenamefont {Chen}}]{minnich2009bulk}%
  \BibitemOpen
  \bibfield  {author} {\bibinfo {author} {\bibfnamefont {A.~J.}\ \bibnamefont
  {Minnich}}, \bibinfo {author} {\bibfnamefont {M.~S.}\ \bibnamefont
  {Dresselhaus}}, \bibinfo {author} {\bibfnamefont {Z.~F.}\ \bibnamefont
  {Ren}},\ and\ \bibinfo {author} {\bibfnamefont {G.}~\bibnamefont {Chen}},\
  }\bibfield  {title} {\bibinfo {title} {Bulk nanostructured thermoelectric
  materials: {Current} research and future prospects},\ }\href
  {https://doi.org/10.1039/b822664b} {\bibfield  {journal} {\bibinfo  {journal}
  {Energy Environ. Sci.}\ }\textbf {\bibinfo {volume} {2}},\ \bibinfo {pages}
  {466} (\bibinfo {year} {2009})}\BibitemShut {NoStop}%
\bibitem [{\citenamefont {Biswas}\ \emph {et~al.}(2012)\citenamefont {Biswas},
  \citenamefont {He}, \citenamefont {Blum}, \citenamefont {Wu}, \citenamefont
  {Hogan}, \citenamefont {Seidman}, \citenamefont {Dravid},\ and\ \citenamefont
  {Kanatzidis}}]{biswas2012high}%
  \BibitemOpen
  \bibfield  {author} {\bibinfo {author} {\bibfnamefont {K.}~\bibnamefont
  {Biswas}}, \bibinfo {author} {\bibfnamefont {J.}~\bibnamefont {He}}, \bibinfo
  {author} {\bibfnamefont {I.~D.}\ \bibnamefont {Blum}}, \bibinfo {author}
  {\bibfnamefont {C.-I.}\ \bibnamefont {Wu}}, \bibinfo {author} {\bibfnamefont
  {T.~P.}\ \bibnamefont {Hogan}}, \bibinfo {author} {\bibfnamefont {D.~N.}\
  \bibnamefont {Seidman}}, \bibinfo {author} {\bibfnamefont {V.~P.}\
  \bibnamefont {Dravid}},\ and\ \bibinfo {author} {\bibfnamefont {M.~G.}\
  \bibnamefont {Kanatzidis}},\ }\bibfield  {title} {\bibinfo {title}
  {High-performance bulk thermoelectrics with all-scale hierarchical
  architectures},\ }\href {https://doi.org/10.1038/nature11439} {\bibfield
  {journal} {\bibinfo  {journal} {Nature}\ }\textbf {\bibinfo {volume} {489}},\
  \bibinfo {pages} {414} (\bibinfo {year} {2012})}\BibitemShut {NoStop}%
\bibitem [{\citenamefont {Poudel}\ \emph {et~al.}(2008)\citenamefont {Poudel},
  \citenamefont {Hao}, \citenamefont {Ma}, \citenamefont {Lan}, \citenamefont
  {Minnich}, \citenamefont {Yu}, \citenamefont {Yan}, \citenamefont {Wang},
  \citenamefont {Muto}, \citenamefont {Vashaee}, \citenamefont {Chen},
  \citenamefont {Liu}, \citenamefont {Dresselhaus}, \citenamefont {Chen},\ and\
  \citenamefont {Ren}}]{poudel2008high}%
  \BibitemOpen
  \bibfield  {author} {\bibinfo {author} {\bibfnamefont {B.}~\bibnamefont
  {Poudel}}, \bibinfo {author} {\bibfnamefont {Q.}~\bibnamefont {Hao}},
  \bibinfo {author} {\bibfnamefont {Y.}~\bibnamefont {Ma}}, \bibinfo {author}
  {\bibfnamefont {Y.}~\bibnamefont {Lan}}, \bibinfo {author} {\bibfnamefont
  {A.}~\bibnamefont {Minnich}}, \bibinfo {author} {\bibfnamefont
  {B.}~\bibnamefont {Yu}}, \bibinfo {author} {\bibfnamefont {X.}~\bibnamefont
  {Yan}}, \bibinfo {author} {\bibfnamefont {D.}~\bibnamefont {Wang}}, \bibinfo
  {author} {\bibfnamefont {A.}~\bibnamefont {Muto}}, \bibinfo {author}
  {\bibfnamefont {D.}~\bibnamefont {Vashaee}}, \bibinfo {author} {\bibfnamefont
  {X.}~\bibnamefont {Chen}}, \bibinfo {author} {\bibfnamefont {J.}~\bibnamefont
  {Liu}}, \bibinfo {author} {\bibfnamefont {M.~S.}\ \bibnamefont
  {Dresselhaus}}, \bibinfo {author} {\bibfnamefont {G.}~\bibnamefont {Chen}},\
  and\ \bibinfo {author} {\bibfnamefont {Z.}~\bibnamefont {Ren}},\ }\bibfield
  {title} {\bibinfo {title} {High-thermoelectric performance of nanostructured
  bismuth antimony {Telluride} bulk alloys},\ }\href
  {https://doi.org/10.1126/science.1156446} {\bibfield  {journal} {\bibinfo
  {journal} {Science}\ }\textbf {\bibinfo {volume} {320}},\ \bibinfo {pages}
  {634} (\bibinfo {year} {2008})}\BibitemShut {NoStop}%
\bibitem [{\citenamefont {Ravichandran}\ \emph {et~al.}(2013)\citenamefont
  {Ravichandran}, \citenamefont {Yadav}, \citenamefont {Cheaito}, \citenamefont
  {Rossen}, \citenamefont {Soukiassian}, \citenamefont {Suresha}, \citenamefont
  {Duda}, \citenamefont {Foley}, \citenamefont {Lee}, \citenamefont {Zhu},
  \citenamefont {Lichtenberger}, \citenamefont {Moore}, \citenamefont {Muller},
  \citenamefont {Schlom}, \citenamefont {Hopkins}, \citenamefont {Majumdar},
  \citenamefont {Ramesh},\ and\ \citenamefont
  {Zurbuchen}}]{ravichandran2014crossover}%
  \BibitemOpen
  \bibfield  {author} {\bibinfo {author} {\bibfnamefont {J.}~\bibnamefont
  {Ravichandran}}, \bibinfo {author} {\bibfnamefont {A.~K.}\ \bibnamefont
  {Yadav}}, \bibinfo {author} {\bibfnamefont {R.}~\bibnamefont {Cheaito}},
  \bibinfo {author} {\bibfnamefont {P.~B.}\ \bibnamefont {Rossen}}, \bibinfo
  {author} {\bibfnamefont {A.}~\bibnamefont {Soukiassian}}, \bibinfo {author}
  {\bibfnamefont {S.~J.}\ \bibnamefont {Suresha}}, \bibinfo {author}
  {\bibfnamefont {J.~C.}\ \bibnamefont {Duda}}, \bibinfo {author}
  {\bibfnamefont {B.~M.}\ \bibnamefont {Foley}}, \bibinfo {author}
  {\bibfnamefont {C.-H.}\ \bibnamefont {Lee}}, \bibinfo {author} {\bibfnamefont
  {Y.}~\bibnamefont {Zhu}}, \bibinfo {author} {\bibfnamefont {A.~W.}\
  \bibnamefont {Lichtenberger}}, \bibinfo {author} {\bibfnamefont {J.~E.}\
  \bibnamefont {Moore}}, \bibinfo {author} {\bibfnamefont {D.~A.}\ \bibnamefont
  {Muller}}, \bibinfo {author} {\bibfnamefont {D.~G.}\ \bibnamefont {Schlom}},
  \bibinfo {author} {\bibfnamefont {P.~E.}\ \bibnamefont {Hopkins}}, \bibinfo
  {author} {\bibfnamefont {A.}~\bibnamefont {Majumdar}}, \bibinfo {author}
  {\bibfnamefont {R.}~\bibnamefont {Ramesh}},\ and\ \bibinfo {author}
  {\bibfnamefont {M.~A.}\ \bibnamefont {Zurbuchen}},\ }\bibfield  {title}
  {\bibinfo {title} {Crossover from incoherent to coherent phonon scattering in
  epitaxial oxide superlattices},\ }\href {https://doi.org/10.1038/nmat3826}
  {\bibfield  {journal} {\bibinfo  {journal} {Nat. Mater.}\ }\textbf {\bibinfo
  {volume} {13}},\ \bibinfo {pages} {168} (\bibinfo {year} {2013})}\BibitemShut
  {NoStop}%
\bibitem [{\citenamefont {Sun}\ \emph {et~al.}(2020)\citenamefont {Sun},
  \citenamefont {Niu}, \citenamefont {Hermann}, \citenamefont {Moon},
  \citenamefont {Shulumba}, \citenamefont {Page}, \citenamefont {Zhao},
  \citenamefont {Thind}, \citenamefont {Mahalingam}, \citenamefont
  {Milam-Guerrero}, \citenamefont {Haiges}, \citenamefont {Mecklenburg},
  \citenamefont {Melot}, \citenamefont {Jho}, \citenamefont {Howe},
  \citenamefont {Mishra}, \citenamefont {Alatas}, \citenamefont {Winn},
  \citenamefont {Manley}, \citenamefont {Ravichandran},\ and\ \citenamefont
  {Minnich}}]{sun2020high}%
  \BibitemOpen
  \bibfield  {author} {\bibinfo {author} {\bibfnamefont {B.}~\bibnamefont
  {Sun}}, \bibinfo {author} {\bibfnamefont {S.}~\bibnamefont {Niu}}, \bibinfo
  {author} {\bibfnamefont {R.~P.}\ \bibnamefont {Hermann}}, \bibinfo {author}
  {\bibfnamefont {J.}~\bibnamefont {Moon}}, \bibinfo {author} {\bibfnamefont
  {N.}~\bibnamefont {Shulumba}}, \bibinfo {author} {\bibfnamefont
  {K.}~\bibnamefont {Page}}, \bibinfo {author} {\bibfnamefont {B.}~\bibnamefont
  {Zhao}}, \bibinfo {author} {\bibfnamefont {A.~S.}\ \bibnamefont {Thind}},
  \bibinfo {author} {\bibfnamefont {K.}~\bibnamefont {Mahalingam}}, \bibinfo
  {author} {\bibfnamefont {J.}~\bibnamefont {Milam-Guerrero}}, \bibinfo
  {author} {\bibfnamefont {R.}~\bibnamefont {Haiges}}, \bibinfo {author}
  {\bibfnamefont {M.}~\bibnamefont {Mecklenburg}}, \bibinfo {author}
  {\bibfnamefont {B.~C.}\ \bibnamefont {Melot}}, \bibinfo {author}
  {\bibfnamefont {Y.-D.}\ \bibnamefont {Jho}}, \bibinfo {author} {\bibfnamefont
  {B.~M.}\ \bibnamefont {Howe}}, \bibinfo {author} {\bibfnamefont
  {R.}~\bibnamefont {Mishra}}, \bibinfo {author} {\bibfnamefont
  {A.}~\bibnamefont {Alatas}}, \bibinfo {author} {\bibfnamefont
  {B.}~\bibnamefont {Winn}}, \bibinfo {author} {\bibfnamefont {M.~E.}\
  \bibnamefont {Manley}}, \bibinfo {author} {\bibfnamefont {J.}~\bibnamefont
  {Ravichandran}},\ and\ \bibinfo {author} {\bibfnamefont {A.~J.}\ \bibnamefont
  {Minnich}},\ }\bibfield  {title} {\bibinfo {title} {High frequency atomic
  tunneling yields ultralow and glass-like thermal conductivity in chalcogenide
  single crystals},\ }\href {https://doi.org/10.1038/s41467-020-19872-w}
  {\bibfield  {journal} {\bibinfo  {journal} {Nat. Commun.}\ }\textbf {\bibinfo
  {volume} {11}},\ \bibinfo {pages} {6039} (\bibinfo {year}
  {2020})}\BibitemShut {NoStop}%
\bibitem [{\citenamefont {Ma}\ \emph {et~al.}(2013)\citenamefont {Ma},
  \citenamefont {Delaire}, \citenamefont {May}, \citenamefont {Carlton},
  \citenamefont {McGuire}, \citenamefont {VanBebber}, \citenamefont
  {Abernathy}, \citenamefont {Ehlers}, \citenamefont {Hong}, \citenamefont
  {Huq}, \citenamefont {Tian}, \citenamefont {Keppens}, \citenamefont
  {Shao-Horn},\ and\ \citenamefont {Sales}}]{ma2013glass}%
  \BibitemOpen
  \bibfield  {author} {\bibinfo {author} {\bibfnamefont {J.}~\bibnamefont
  {Ma}}, \bibinfo {author} {\bibfnamefont {O.}~\bibnamefont {Delaire}},
  \bibinfo {author} {\bibfnamefont {A.~F.}\ \bibnamefont {May}}, \bibinfo
  {author} {\bibfnamefont {C.~E.}\ \bibnamefont {Carlton}}, \bibinfo {author}
  {\bibfnamefont {M.~A.}\ \bibnamefont {McGuire}}, \bibinfo {author}
  {\bibfnamefont {L.~H.}\ \bibnamefont {VanBebber}}, \bibinfo {author}
  {\bibfnamefont {D.~L.}\ \bibnamefont {Abernathy}}, \bibinfo {author}
  {\bibfnamefont {G.}~\bibnamefont {Ehlers}}, \bibinfo {author} {\bibfnamefont
  {T.}~\bibnamefont {Hong}}, \bibinfo {author} {\bibfnamefont {A.}~\bibnamefont
  {Huq}}, \bibinfo {author} {\bibfnamefont {W.}~\bibnamefont {Tian}}, \bibinfo
  {author} {\bibfnamefont {V.~M.}\ \bibnamefont {Keppens}}, \bibinfo {author}
  {\bibfnamefont {Y.}~\bibnamefont {Shao-Horn}},\ and\ \bibinfo {author}
  {\bibfnamefont {B.~C.}\ \bibnamefont {Sales}},\ }\bibfield  {title} {\bibinfo
  {title} {Glass-like phonon scattering from a spontaneous nanostructure in
  {AgSbTe$_2$}},\ }\href {https://doi.org/10.1038/nnano.2013.95} {\bibfield
  {journal} {\bibinfo  {journal} {Nat. Nanotechnol.}\ }\textbf {\bibinfo
  {volume} {8}},\ \bibinfo {pages} {445} (\bibinfo {year} {2013})}\BibitemShut
  {NoStop}%
\bibitem [{\citenamefont {Zhao}\ \emph {et~al.}(2014)\citenamefont {Zhao},
  \citenamefont {Lo}, \citenamefont {Zhang}, \citenamefont {Sun}, \citenamefont
  {Tan}, \citenamefont {Uher}, \citenamefont {Wolverton}, \citenamefont
  {Dravid},\ and\ \citenamefont {Kanatzidis}}]{zhao2014ultralow}%
  \BibitemOpen
  \bibfield  {author} {\bibinfo {author} {\bibfnamefont {L.-D.}\ \bibnamefont
  {Zhao}}, \bibinfo {author} {\bibfnamefont {S.-H.}\ \bibnamefont {Lo}},
  \bibinfo {author} {\bibfnamefont {Y.}~\bibnamefont {Zhang}}, \bibinfo
  {author} {\bibfnamefont {H.}~\bibnamefont {Sun}}, \bibinfo {author}
  {\bibfnamefont {G.}~\bibnamefont {Tan}}, \bibinfo {author} {\bibfnamefont
  {C.}~\bibnamefont {Uher}}, \bibinfo {author} {\bibfnamefont {C.}~\bibnamefont
  {Wolverton}}, \bibinfo {author} {\bibfnamefont {V.~P.}\ \bibnamefont
  {Dravid}},\ and\ \bibinfo {author} {\bibfnamefont {M.~G.}\ \bibnamefont
  {Kanatzidis}},\ }\bibfield  {title} {\bibinfo {title} {Ultralow thermal
  conductivity and high thermoelectric figure of merit in {SnSe} crystals},\
  }\href {https://doi.org/10.1038/nature13184} {\bibfield  {journal} {\bibinfo
  {journal} {Nature}\ }\textbf {\bibinfo {volume} {508}},\ \bibinfo {pages}
  {373} (\bibinfo {year} {2014})}\BibitemShut {NoStop}%
\bibitem [{\citenamefont {Lu}\ \emph {et~al.}(2012)\citenamefont {Lu},
  \citenamefont {Morelli}, \citenamefont {Xia}, \citenamefont {Zhou},
  \citenamefont {Ozolins}, \citenamefont {Chi}, \citenamefont {Zhou},\ and\
  \citenamefont {Uher}}]{lu2013high}%
  \BibitemOpen
  \bibfield  {author} {\bibinfo {author} {\bibfnamefont {X.}~\bibnamefont
  {Lu}}, \bibinfo {author} {\bibfnamefont {D.~T.}\ \bibnamefont {Morelli}},
  \bibinfo {author} {\bibfnamefont {Y.}~\bibnamefont {Xia}}, \bibinfo {author}
  {\bibfnamefont {F.}~\bibnamefont {Zhou}}, \bibinfo {author} {\bibfnamefont
  {V.}~\bibnamefont {Ozolins}}, \bibinfo {author} {\bibfnamefont
  {H.}~\bibnamefont {Chi}}, \bibinfo {author} {\bibfnamefont {X.}~\bibnamefont
  {Zhou}},\ and\ \bibinfo {author} {\bibfnamefont {C.}~\bibnamefont {Uher}},\
  }\bibfield  {title} {\bibinfo {title} {High performance thermoelectricity in
  earth-abundant compounds based on natural mineral tetrahedrites},\ }\href
  {https://doi.org/10.1002/aenm.201200650} {\bibfield  {journal} {\bibinfo
  {journal} {Adv. Energy Mater.}\ }\textbf {\bibinfo {volume} {3}},\ \bibinfo
  {pages} {342} (\bibinfo {year} {2012})}\BibitemShut {NoStop}%
\bibitem [{\citenamefont {He}\ \emph {et~al.}(2021)\citenamefont {He},
  \citenamefont {Xia}, \citenamefont {Lin}, \citenamefont {Pal}, \citenamefont
  {Zhu}, \citenamefont {Kanatzidis},\ and\ \citenamefont
  {Wolverton}}]{he2022accelerated}%
  \BibitemOpen
  \bibfield  {author} {\bibinfo {author} {\bibfnamefont {J.}~\bibnamefont
  {He}}, \bibinfo {author} {\bibfnamefont {Y.}~\bibnamefont {Xia}}, \bibinfo
  {author} {\bibfnamefont {W.}~\bibnamefont {Lin}}, \bibinfo {author}
  {\bibfnamefont {K.}~\bibnamefont {Pal}}, \bibinfo {author} {\bibfnamefont
  {Y.}~\bibnamefont {Zhu}}, \bibinfo {author} {\bibfnamefont {M.~G.}\
  \bibnamefont {Kanatzidis}},\ and\ \bibinfo {author} {\bibfnamefont
  {C.}~\bibnamefont {Wolverton}},\ }\bibfield  {title} {\bibinfo {title}
  {Accelerated discovery and design of ultralow lattice thermal conductivity
  materials using chemical bonding principles},\ }\href
  {https://doi.org/10.1002/adfm.202108532} {\bibfield  {journal} {\bibinfo
  {journal} {Adv. Funct. Mater.}\ }\textbf {\bibinfo {volume} {32}},\ \bibinfo
  {pages} {2108532} (\bibinfo {year} {2021})}\BibitemShut {NoStop}%
\bibitem [{\citenamefont {Chiritescu}\ \emph {et~al.}(2007)\citenamefont
  {Chiritescu}, \citenamefont {Cahill}, \citenamefont {Nguyen}, \citenamefont
  {Johnson}, \citenamefont {Bodapati}, \citenamefont {Keblinski},\ and\
  \citenamefont {Zschack}}]{chiritescu2007ultralow}%
  \BibitemOpen
  \bibfield  {author} {\bibinfo {author} {\bibfnamefont {C.}~\bibnamefont
  {Chiritescu}}, \bibinfo {author} {\bibfnamefont {D.~G.}\ \bibnamefont
  {Cahill}}, \bibinfo {author} {\bibfnamefont {N.}~\bibnamefont {Nguyen}},
  \bibinfo {author} {\bibfnamefont {D.}~\bibnamefont {Johnson}}, \bibinfo
  {author} {\bibfnamefont {A.}~\bibnamefont {Bodapati}}, \bibinfo {author}
  {\bibfnamefont {P.}~\bibnamefont {Keblinski}},\ and\ \bibinfo {author}
  {\bibfnamefont {P.}~\bibnamefont {Zschack}},\ }\bibfield  {title} {\bibinfo
  {title} {Ultralow thermal conductivity in disordered, layered {WSe$_2$}
  crystals},\ }\href {https://doi.org/10.1126/science.1136494} {\bibfield
  {journal} {\bibinfo  {journal} {Science}\ }\textbf {\bibinfo {volume}
  {315}},\ \bibinfo {pages} {351} (\bibinfo {year} {2007})}\BibitemShut
  {NoStop}%
\bibitem [{\citenamefont {Costescu}\ \emph {et~al.}(2004)\citenamefont
  {Costescu}, \citenamefont {Cahill}, \citenamefont {Fabreguette},
  \citenamefont {Sechrist},\ and\ \citenamefont {George}}]{costescu2004ultra}%
  \BibitemOpen
  \bibfield  {author} {\bibinfo {author} {\bibfnamefont {R.~M.}\ \bibnamefont
  {Costescu}}, \bibinfo {author} {\bibfnamefont {D.~G.}\ \bibnamefont
  {Cahill}}, \bibinfo {author} {\bibfnamefont {F.~H.}\ \bibnamefont
  {Fabreguette}}, \bibinfo {author} {\bibfnamefont {Z.~A.}\ \bibnamefont
  {Sechrist}},\ and\ \bibinfo {author} {\bibfnamefont {S.~M.}\ \bibnamefont
  {George}},\ }\bibfield  {title} {\bibinfo {title} {Ultra-low thermal
  conductivity in {W/Al$_2$O$_3$} nanolaminates},\ }\href
  {https://doi.org/10.1126/science.1093711} {\bibfield  {journal} {\bibinfo
  {journal} {Science}\ }\textbf {\bibinfo {volume} {303}},\ \bibinfo {pages}
  {989} (\bibinfo {year} {2004})}\BibitemShut {NoStop}%
\bibitem [{\citenamefont {Liu}\ \emph {et~al.}(2012)\citenamefont {Liu},
  \citenamefont {Shi}, \citenamefont {Xu}, \citenamefont {Zhang}, \citenamefont
  {Zhang}, \citenamefont {Chen}, \citenamefont {Li}, \citenamefont {Uher},
  \citenamefont {Day},\ and\ \citenamefont {Snyder}}]{liu2012copper}%
  \BibitemOpen
  \bibfield  {author} {\bibinfo {author} {\bibfnamefont {H.}~\bibnamefont
  {Liu}}, \bibinfo {author} {\bibfnamefont {X.}~\bibnamefont {Shi}}, \bibinfo
  {author} {\bibfnamefont {F.}~\bibnamefont {Xu}}, \bibinfo {author}
  {\bibfnamefont {L.}~\bibnamefont {Zhang}}, \bibinfo {author} {\bibfnamefont
  {W.}~\bibnamefont {Zhang}}, \bibinfo {author} {\bibfnamefont
  {L.}~\bibnamefont {Chen}}, \bibinfo {author} {\bibfnamefont {Q.}~\bibnamefont
  {Li}}, \bibinfo {author} {\bibfnamefont {C.}~\bibnamefont {Uher}}, \bibinfo
  {author} {\bibfnamefont {T.}~\bibnamefont {Day}},\ and\ \bibinfo {author}
  {\bibfnamefont {G.~J.}\ \bibnamefont {Snyder}},\ }\bibfield  {title}
  {\bibinfo {title} {Copper ion liquid-like thermoelectrics},\ }\href
  {https://doi.org/10.1038/nmat3273} {\bibfield  {journal} {\bibinfo  {journal}
  {Nat. Mater.}\ }\textbf {\bibinfo {volume} {11}},\ \bibinfo {pages} {422}
  (\bibinfo {year} {2012})}\BibitemShut {NoStop}%
\bibitem [{\citenamefont {Roychowdhury}\ \emph {et~al.}(2018)\citenamefont
  {Roychowdhury}, \citenamefont {Jana}, \citenamefont {Pan}, \citenamefont
  {Guin}, \citenamefont {Sanyal}, \citenamefont {Waghmare},\ and\ \citenamefont
  {Biswas}}]{roychowdhury2018soft}%
  \BibitemOpen
  \bibfield  {author} {\bibinfo {author} {\bibfnamefont {S.}~\bibnamefont
  {Roychowdhury}}, \bibinfo {author} {\bibfnamefont {M.~K.}\ \bibnamefont
  {Jana}}, \bibinfo {author} {\bibfnamefont {J.}~\bibnamefont {Pan}}, \bibinfo
  {author} {\bibfnamefont {S.~N.}\ \bibnamefont {Guin}}, \bibinfo {author}
  {\bibfnamefont {D.}~\bibnamefont {Sanyal}}, \bibinfo {author} {\bibfnamefont
  {U.~V.}\ \bibnamefont {Waghmare}},\ and\ \bibinfo {author} {\bibfnamefont
  {K.}~\bibnamefont {Biswas}},\ }\bibfield  {title} {\bibinfo {title} {Soft
  phonon modes leading to ultralow thermal conductivity and high thermoelectric
  performance in {AgCuTe}},\ }\href {https://doi.org/10.1002/ange.201801491}
  {\bibfield  {journal} {\bibinfo  {journal} {Angew. Chem.}\ }\textbf {\bibinfo
  {volume} {130}},\ \bibinfo {pages} {4107} (\bibinfo {year}
  {2018})}\BibitemShut {NoStop}%
\bibitem [{\citenamefont {Li}\ \emph {et~al.}(2018{\natexlab{a}})\citenamefont
  {Li}, \citenamefont {Wang}, \citenamefont {Kawakita}, \citenamefont {Zhang},
  \citenamefont {Feygenson}, \citenamefont {Yu}, \citenamefont {Wu},
  \citenamefont {Ohara}, \citenamefont {Kikuchi}, \citenamefont {Shibata},
  \citenamefont {Yamada}, \citenamefont {Ning}, \citenamefont {Chen},
  \citenamefont {He}, \citenamefont {Vaknin}, \citenamefont {Wu}, \citenamefont
  {Nakajima},\ and\ \citenamefont {Kanatzidis}}]{li2018liquid}%
  \BibitemOpen
  \bibfield  {author} {\bibinfo {author} {\bibfnamefont {B.}~\bibnamefont
  {Li}}, \bibinfo {author} {\bibfnamefont {H.}~\bibnamefont {Wang}}, \bibinfo
  {author} {\bibfnamefont {Y.}~\bibnamefont {Kawakita}}, \bibinfo {author}
  {\bibfnamefont {Q.}~\bibnamefont {Zhang}}, \bibinfo {author} {\bibfnamefont
  {M.}~\bibnamefont {Feygenson}}, \bibinfo {author} {\bibfnamefont {H.~L.}\
  \bibnamefont {Yu}}, \bibinfo {author} {\bibfnamefont {D.}~\bibnamefont {Wu}},
  \bibinfo {author} {\bibfnamefont {K.}~\bibnamefont {Ohara}}, \bibinfo
  {author} {\bibfnamefont {T.}~\bibnamefont {Kikuchi}}, \bibinfo {author}
  {\bibfnamefont {K.}~\bibnamefont {Shibata}}, \bibinfo {author} {\bibfnamefont
  {T.}~\bibnamefont {Yamada}}, \bibinfo {author} {\bibfnamefont {X.~K.}\
  \bibnamefont {Ning}}, \bibinfo {author} {\bibfnamefont {Y.}~\bibnamefont
  {Chen}}, \bibinfo {author} {\bibfnamefont {J.~Q.}\ \bibnamefont {He}},
  \bibinfo {author} {\bibfnamefont {D.}~\bibnamefont {Vaknin}}, \bibinfo
  {author} {\bibfnamefont {R.~Q.}\ \bibnamefont {Wu}}, \bibinfo {author}
  {\bibfnamefont {K.}~\bibnamefont {Nakajima}},\ and\ \bibinfo {author}
  {\bibfnamefont {M.~G.}\ \bibnamefont {Kanatzidis}},\ }\bibfield  {title}
  {\bibinfo {title} {Liquid-like thermal conduction in intercalated layered
  crystalline solids},\ }\href {https://doi.org/10.1038/s41563-017-0004-2}
  {\bibfield  {journal} {\bibinfo  {journal} {Nat. Mater.}\ }\textbf {\bibinfo
  {volume} {17}},\ \bibinfo {pages} {226} (\bibinfo {year}
  {2018}{\natexlab{a}})}\BibitemShut {NoStop}%
\bibitem [{\citenamefont {Xie}\ \emph {et~al.}(2020)\citenamefont {Xie},
  \citenamefont {Hao}, \citenamefont {Bao}, \citenamefont {Slade},
  \citenamefont {Snyder}, \citenamefont {Wolverton},\ and\ \citenamefont
  {Kanatzidis}}]{xie2020all}%
  \BibitemOpen
  \bibfield  {author} {\bibinfo {author} {\bibfnamefont {H.}~\bibnamefont
  {Xie}}, \bibinfo {author} {\bibfnamefont {S.}~\bibnamefont {Hao}}, \bibinfo
  {author} {\bibfnamefont {J.}~\bibnamefont {Bao}}, \bibinfo {author}
  {\bibfnamefont {T.~J.}\ \bibnamefont {Slade}}, \bibinfo {author}
  {\bibfnamefont {G.~J.}\ \bibnamefont {Snyder}}, \bibinfo {author}
  {\bibfnamefont {C.}~\bibnamefont {Wolverton}},\ and\ \bibinfo {author}
  {\bibfnamefont {M.~G.}\ \bibnamefont {Kanatzidis}},\ }\bibfield  {title}
  {\bibinfo {title} {All-inorganic halide perovskites as potential
  thermoelectric materials: {Dynamic} cation off-centering induces ultralow
  thermal conductivity},\ }\href {https://doi.org/10.1021/jacs.0c03427}
  {\bibfield  {journal} {\bibinfo  {journal} {JACS}\ }\textbf {\bibinfo
  {volume} {142}},\ \bibinfo {pages} {9553} (\bibinfo {year}
  {2020})}\BibitemShut {NoStop}%
\bibitem [{\citenamefont {Delaire}\ \emph {et~al.}(2011)\citenamefont
  {Delaire}, \citenamefont {Ma}, \citenamefont {Marty}, \citenamefont {May},
  \citenamefont {McGuire}, \citenamefont {Du}, \citenamefont {Singh},
  \citenamefont {Podlesnyak}, \citenamefont {Ehlers}, \citenamefont {Lumsden},\
  and\ \citenamefont {Sales}}]{delaire2011giant}%
  \BibitemOpen
  \bibfield  {author} {\bibinfo {author} {\bibfnamefont {O.}~\bibnamefont
  {Delaire}}, \bibinfo {author} {\bibfnamefont {J.}~\bibnamefont {Ma}},
  \bibinfo {author} {\bibfnamefont {K.}~\bibnamefont {Marty}}, \bibinfo
  {author} {\bibfnamefont {A.~F.}\ \bibnamefont {May}}, \bibinfo {author}
  {\bibfnamefont {M.~A.}\ \bibnamefont {McGuire}}, \bibinfo {author}
  {\bibfnamefont {M.-H.}\ \bibnamefont {Du}}, \bibinfo {author} {\bibfnamefont
  {D.~J.}\ \bibnamefont {Singh}}, \bibinfo {author} {\bibfnamefont
  {A.}~\bibnamefont {Podlesnyak}}, \bibinfo {author} {\bibfnamefont
  {G.}~\bibnamefont {Ehlers}}, \bibinfo {author} {\bibfnamefont {M.~D.}\
  \bibnamefont {Lumsden}},\ and\ \bibinfo {author} {\bibfnamefont {B.~C.}\
  \bibnamefont {Sales}},\ }\bibfield  {title} {\bibinfo {title} {Giant
  anharmonic phonon scattering in {PbTe}},\ }\href
  {https://doi.org/10.1038/nmat3035} {\bibfield  {journal} {\bibinfo  {journal}
  {Nat. Mater.}\ }\textbf {\bibinfo {volume} {10}},\ \bibinfo {pages} {614}
  (\bibinfo {year} {2011})}\BibitemShut {NoStop}%
\bibitem [{\citenamefont {Sarkar}\ \emph {et~al.}(2020)\citenamefont {Sarkar},
  \citenamefont {Ghosh}, \citenamefont {Roychowdhury}, \citenamefont {Arora},
  \citenamefont {Sajan}, \citenamefont {Sheet}, \citenamefont {Waghmare},\ and\
  \citenamefont {Biswas}}]{sarkar2020ferroelectric}%
  \BibitemOpen
  \bibfield  {author} {\bibinfo {author} {\bibfnamefont {D.}~\bibnamefont
  {Sarkar}}, \bibinfo {author} {\bibfnamefont {T.}~\bibnamefont {Ghosh}},
  \bibinfo {author} {\bibfnamefont {S.}~\bibnamefont {Roychowdhury}}, \bibinfo
  {author} {\bibfnamefont {R.}~\bibnamefont {Arora}}, \bibinfo {author}
  {\bibfnamefont {S.}~\bibnamefont {Sajan}}, \bibinfo {author} {\bibfnamefont
  {G.}~\bibnamefont {Sheet}}, \bibinfo {author} {\bibfnamefont {U.~V.}\
  \bibnamefont {Waghmare}},\ and\ \bibinfo {author} {\bibfnamefont
  {K.}~\bibnamefont {Biswas}},\ }\bibfield  {title} {\bibinfo {title}
  {Ferroelectric instability induced ultralow thermal conductivity and high
  thermoelectric performance in rhombohedral $p$-type {GeSe} crystal},\ }\href
  {https://doi.org/10.1021/jacs.0c03696} {\bibfield  {journal} {\bibinfo
  {journal} {JACS}\ }\textbf {\bibinfo {volume} {142}},\ \bibinfo {pages}
  {12237} (\bibinfo {year} {2020})}\BibitemShut {NoStop}%
\bibitem [{\citenamefont {Tritt}(2001)}]{tritt2001recent}%
  \BibitemOpen
  \bibfield  {author} {\bibinfo {author} {\bibfnamefont {T.}~\bibnamefont
  {Tritt}},\ }\href
  {https://www.elsevier.com/books/recent-trends-in-thermoelectric-materials-research-part-three/tritt/978-0-12-752180-0}
  {\emph {\bibinfo {title} {Recent trends in thermoelectric materials research:
  {Part} three}}}\ (\bibinfo  {publisher} {Elsevier},\ \bibinfo {year}
  {2001})\BibitemShut {NoStop}%
\bibitem [{\citenamefont {Tadano}\ \emph {et~al.}(2015)\citenamefont {Tadano},
  \citenamefont {Gohda},\ and\ \citenamefont {Tsuneyuki}}]{tadano2015impact}%
  \BibitemOpen
  \bibfield  {author} {\bibinfo {author} {\bibfnamefont {T.}~\bibnamefont
  {Tadano}}, \bibinfo {author} {\bibfnamefont {Y.}~\bibnamefont {Gohda}},\ and\
  \bibinfo {author} {\bibfnamefont {S.}~\bibnamefont {Tsuneyuki}},\ }\bibfield
  {title} {\bibinfo {title} {Impact of rattlers on thermal conductivity of a
  thermoelectric clathrate: {A} first-principles study},\ }\href
  {https://doi.org/10.1103/physrevlett.114.095501} {\bibfield  {journal}
  {\bibinfo  {journal} {Phys. Rev. Lett.}\ }\textbf {\bibinfo {volume} {114}},\
  \bibinfo {pages} {095501} (\bibinfo {year} {2015})}\BibitemShut {NoStop}%
\bibitem [{\citenamefont {Lin}\ \emph {et~al.}(2016)\citenamefont {Lin},
  \citenamefont {Tan}, \citenamefont {Shen}, \citenamefont {Hao}, \citenamefont
  {Wu}, \citenamefont {Calta}, \citenamefont {Malliakas}, \citenamefont {Wang},
  \citenamefont {Uher}, \citenamefont {Wolverton},\ and\ \citenamefont
  {Kanatzidis}}]{lin2016concerted}%
  \BibitemOpen
  \bibfield  {author} {\bibinfo {author} {\bibfnamefont {H.}~\bibnamefont
  {Lin}}, \bibinfo {author} {\bibfnamefont {G.}~\bibnamefont {Tan}}, \bibinfo
  {author} {\bibfnamefont {J.-N.}\ \bibnamefont {Shen}}, \bibinfo {author}
  {\bibfnamefont {S.}~\bibnamefont {Hao}}, \bibinfo {author} {\bibfnamefont
  {L.-M.}\ \bibnamefont {Wu}}, \bibinfo {author} {\bibfnamefont
  {N.}~\bibnamefont {Calta}}, \bibinfo {author} {\bibfnamefont
  {C.}~\bibnamefont {Malliakas}}, \bibinfo {author} {\bibfnamefont
  {S.}~\bibnamefont {Wang}}, \bibinfo {author} {\bibfnamefont {C.}~\bibnamefont
  {Uher}}, \bibinfo {author} {\bibfnamefont {C.}~\bibnamefont {Wolverton}},\
  and\ \bibinfo {author} {\bibfnamefont {M.~G.}\ \bibnamefont {Kanatzidis}},\
  }\bibfield  {title} {\bibinfo {title} {Concerted rattling in {CsAg$_5$Te$_3$}
  leading to ultralow thermal conductivity and high thermoelectric
  performance},\ }\href {https://doi.org/10.1002/anie.201605015} {\bibfield
  {journal} {\bibinfo  {journal} {Angew. Chem. Int. Ed.}\ }\textbf {\bibinfo
  {volume} {55}},\ \bibinfo {pages} {11431} (\bibinfo {year}
  {2016})}\BibitemShut {NoStop}%
\bibitem [{\citenamefont {He}\ \emph {et~al.}(2016)\citenamefont {He},
  \citenamefont {Amsler}, \citenamefont {Xia}, \citenamefont {Naghavi},
  \citenamefont {Hegde}, \citenamefont {Hao}, \citenamefont {Goedecker},
  \citenamefont {Ozoli\c{n}\v{s}},\ and\ \citenamefont
  {Wolverton}}]{he2016ultralow}%
  \BibitemOpen
  \bibfield  {author} {\bibinfo {author} {\bibfnamefont {J.}~\bibnamefont
  {He}}, \bibinfo {author} {\bibfnamefont {M.}~\bibnamefont {Amsler}}, \bibinfo
  {author} {\bibfnamefont {Y.}~\bibnamefont {Xia}}, \bibinfo {author}
  {\bibfnamefont {S.~S.}\ \bibnamefont {Naghavi}}, \bibinfo {author}
  {\bibfnamefont {V.~I.}\ \bibnamefont {Hegde}}, \bibinfo {author}
  {\bibfnamefont {S.}~\bibnamefont {Hao}}, \bibinfo {author} {\bibfnamefont
  {S.}~\bibnamefont {Goedecker}}, \bibinfo {author} {\bibfnamefont
  {V.}~\bibnamefont {Ozoli\c{n}\v{s}}},\ and\ \bibinfo {author} {\bibfnamefont
  {C.}~\bibnamefont {Wolverton}},\ }\bibfield  {title} {\bibinfo {title}
  {Ultralow thermal conductivity in full heusler semiconductors},\ }\href
  {https://doi.org/10.1103/physrevlett.117.046602} {\bibfield  {journal}
  {\bibinfo  {journal} {Phys. Rev. Lett.}\ }\textbf {\bibinfo {volume} {117}},\
  \bibinfo {pages} {046602} (\bibinfo {year} {2016})}\BibitemShut {NoStop}%
\bibitem [{\citenamefont {Nielsen}\ \emph {et~al.}(2013)\citenamefont
  {Nielsen}, \citenamefont {Ozolins},\ and\ \citenamefont
  {Heremans}}]{nielsen2013lone}%
  \BibitemOpen
  \bibfield  {author} {\bibinfo {author} {\bibfnamefont {M.~D.}\ \bibnamefont
  {Nielsen}}, \bibinfo {author} {\bibfnamefont {V.}~\bibnamefont {Ozolins}},\
  and\ \bibinfo {author} {\bibfnamefont {J.~P.}\ \bibnamefont {Heremans}},\
  }\bibfield  {title} {\bibinfo {title} {Lone pair electrons minimize lattice
  thermal conductivity},\ }\href {https://doi.org/10.1039/c2ee23391f}
  {\bibfield  {journal} {\bibinfo  {journal} {Energy Environ. Sci.}\ }\textbf
  {\bibinfo {volume} {6}},\ \bibinfo {pages} {570} (\bibinfo {year}
  {2013})}\BibitemShut {NoStop}%
\bibitem [{\citenamefont {Skoug}\ and\ \citenamefont
  {Morelli}(2011)}]{skoug2011role}%
  \BibitemOpen
  \bibfield  {author} {\bibinfo {author} {\bibfnamefont {E.~J.}\ \bibnamefont
  {Skoug}}\ and\ \bibinfo {author} {\bibfnamefont {D.~T.}\ \bibnamefont
  {Morelli}},\ }\bibfield  {title} {\bibinfo {title} {Role of lone-pair
  electrons in producing minimum thermal conductivity in nitrogen-group
  chalcogenide compounds},\ }\href
  {https://doi.org/10.1103/physrevlett.107.235901} {\bibfield  {journal}
  {\bibinfo  {journal} {Phys. Rev. Lett.}\ }\textbf {\bibinfo {volume} {107}},\
  \bibinfo {pages} {235901} (\bibinfo {year} {2011})}\BibitemShut {NoStop}%
\bibitem [{\citenamefont {Dutta}\ \emph {et~al.}(2019)\citenamefont {Dutta},
  \citenamefont {Pal}, \citenamefont {Waghmare},\ and\ \citenamefont
  {Biswas}}]{dutta2019bonding}%
  \BibitemOpen
  \bibfield  {author} {\bibinfo {author} {\bibfnamefont {M.}~\bibnamefont
  {Dutta}}, \bibinfo {author} {\bibfnamefont {K.}~\bibnamefont {Pal}}, \bibinfo
  {author} {\bibfnamefont {U.~V.}\ \bibnamefont {Waghmare}},\ and\ \bibinfo
  {author} {\bibfnamefont {K.}~\bibnamefont {Biswas}},\ }\bibfield  {title}
  {\bibinfo {title} {Bonding heterogeneity and lone pair induced anharmonicity
  resulted in ultralow thermal conductivity and promising thermoelectric
  properties in $n$-type {AgPbBiSe$_3$}},\ }\href
  {https://doi.org/10.1039/c9sc00485h} {\bibfield  {journal} {\bibinfo
  {journal} {Chem. Sci.}\ }\textbf {\bibinfo {volume} {10}},\ \bibinfo {pages}
  {4905} (\bibinfo {year} {2019})}\BibitemShut {NoStop}%
\bibitem [{\citenamefont {Nolas}\ \emph
  {et~al.}(2001{\natexlab{b}})\citenamefont {Nolas}, \citenamefont {Sharp},\
  and\ \citenamefont {Goldsmid}}]{nolas2001thermoelectrics}%
  \BibitemOpen
  \bibfield  {author} {\bibinfo {author} {\bibfnamefont {G.~S.}\ \bibnamefont
  {Nolas}}, \bibinfo {author} {\bibfnamefont {J.}~\bibnamefont {Sharp}},\ and\
  \bibinfo {author} {\bibfnamefont {H.~J.}\ \bibnamefont {Goldsmid}},\ }\href
  {https://doi.org/10.1007/978-3-662-04569-5} {\emph {\bibinfo {title}
  {Thermoelectrics: Basic Principles and New Materials Developments}}},\
  Vol.~\bibinfo {volume} {45}\ (\bibinfo  {publisher} {Springer Berlin
  Heidelberg},\ \bibinfo {year} {2001})\BibitemShut {NoStop}%
\bibitem [{\citenamefont {Chen}\ \emph {et~al.}(2017)\citenamefont {Chen},
  \citenamefont {Jian}, \citenamefont {Li}, \citenamefont {Chang},
  \citenamefont {Ge}, \citenamefont {Hanus}, \citenamefont {Yang},
  \citenamefont {Chen}, \citenamefont {Huang}, \citenamefont {Snyder},\ and\
  \citenamefont {Pei}}]{chen2017lattice}%
  \BibitemOpen
  \bibfield  {author} {\bibinfo {author} {\bibfnamefont {Z.}~\bibnamefont
  {Chen}}, \bibinfo {author} {\bibfnamefont {Z.}~\bibnamefont {Jian}}, \bibinfo
  {author} {\bibfnamefont {W.}~\bibnamefont {Li}}, \bibinfo {author}
  {\bibfnamefont {Y.}~\bibnamefont {Chang}}, \bibinfo {author} {\bibfnamefont
  {B.}~\bibnamefont {Ge}}, \bibinfo {author} {\bibfnamefont {R.}~\bibnamefont
  {Hanus}}, \bibinfo {author} {\bibfnamefont {J.}~\bibnamefont {Yang}},
  \bibinfo {author} {\bibfnamefont {Y.}~\bibnamefont {Chen}}, \bibinfo {author}
  {\bibfnamefont {M.}~\bibnamefont {Huang}}, \bibinfo {author} {\bibfnamefont
  {G.~J.}\ \bibnamefont {Snyder}},\ and\ \bibinfo {author} {\bibfnamefont
  {Y.}~\bibnamefont {Pei}},\ }\bibfield  {title} {\bibinfo {title} {Lattice
  dislocations enhancing thermoelectric {PbTe} in addition to band
  convergence},\ }\href {https://doi.org/10.1002/adma.201606768} {\bibfield
  {journal} {\bibinfo  {journal} {Adv. Mater.}\ }\textbf {\bibinfo {volume}
  {29}},\ \bibinfo {pages} {1606768} (\bibinfo {year} {2017})}\BibitemShut
  {NoStop}%
\bibitem [{\citenamefont {Zhao}\ \emph {et~al.}(2016)\citenamefont {Zhao},
  \citenamefont {Tan}, \citenamefont {Hao}, \citenamefont {He}, \citenamefont
  {Pei}, \citenamefont {Chi}, \citenamefont {Wang}, \citenamefont {Gong},
  \citenamefont {Xu}, \citenamefont {Dravid}, \citenamefont {Uher},
  \citenamefont {Snyder}, \citenamefont {Wolverton},\ and\ \citenamefont
  {Kanatzidis}}]{zhao2016ultrahigh}%
  \BibitemOpen
  \bibfield  {author} {\bibinfo {author} {\bibfnamefont {L.-D.}\ \bibnamefont
  {Zhao}}, \bibinfo {author} {\bibfnamefont {G.}~\bibnamefont {Tan}}, \bibinfo
  {author} {\bibfnamefont {S.}~\bibnamefont {Hao}}, \bibinfo {author}
  {\bibfnamefont {J.}~\bibnamefont {He}}, \bibinfo {author} {\bibfnamefont
  {Y.}~\bibnamefont {Pei}}, \bibinfo {author} {\bibfnamefont {H.}~\bibnamefont
  {Chi}}, \bibinfo {author} {\bibfnamefont {H.}~\bibnamefont {Wang}}, \bibinfo
  {author} {\bibfnamefont {S.}~\bibnamefont {Gong}}, \bibinfo {author}
  {\bibfnamefont {H.}~\bibnamefont {Xu}}, \bibinfo {author} {\bibfnamefont
  {V.~P.}\ \bibnamefont {Dravid}}, \bibinfo {author} {\bibfnamefont
  {C.}~\bibnamefont {Uher}}, \bibinfo {author} {\bibfnamefont {G.~J.}\
  \bibnamefont {Snyder}}, \bibinfo {author} {\bibfnamefont {C.}~\bibnamefont
  {Wolverton}},\ and\ \bibinfo {author} {\bibfnamefont {M.~G.}\ \bibnamefont
  {Kanatzidis}},\ }\bibfield  {title} {\bibinfo {title} {Ultrahigh power factor
  and thermoelectric performance in hole-doped single-crystal {SnSe}},\ }\href
  {https://doi.org/10.1126/science.aad3749} {\bibfield  {journal} {\bibinfo
  {journal} {Science}\ }\textbf {\bibinfo {volume} {351}},\ \bibinfo {pages}
  {141} (\bibinfo {year} {2016})}\BibitemShut {NoStop}%
\bibitem [{\citenamefont {Li}\ \emph {et~al.}(2018{\natexlab{b}})\citenamefont
  {Li}, \citenamefont {Zhang}, \citenamefont {Chen}, \citenamefont {Lin},
  \citenamefont {Li}, \citenamefont {Shen}, \citenamefont {Witting},
  \citenamefont {Faghaninia}, \citenamefont {Chen}, \citenamefont {Jain},
  \citenamefont {Chen}, \citenamefont {Snyder},\ and\ \citenamefont
  {Pei}}]{li2018low}%
  \BibitemOpen
  \bibfield  {author} {\bibinfo {author} {\bibfnamefont {J.}~\bibnamefont
  {Li}}, \bibinfo {author} {\bibfnamefont {X.}~\bibnamefont {Zhang}}, \bibinfo
  {author} {\bibfnamefont {Z.}~\bibnamefont {Chen}}, \bibinfo {author}
  {\bibfnamefont {S.}~\bibnamefont {Lin}}, \bibinfo {author} {\bibfnamefont
  {W.}~\bibnamefont {Li}}, \bibinfo {author} {\bibfnamefont {J.}~\bibnamefont
  {Shen}}, \bibinfo {author} {\bibfnamefont {I.~T.}\ \bibnamefont {Witting}},
  \bibinfo {author} {\bibfnamefont {A.}~\bibnamefont {Faghaninia}}, \bibinfo
  {author} {\bibfnamefont {Y.}~\bibnamefont {Chen}}, \bibinfo {author}
  {\bibfnamefont {A.}~\bibnamefont {Jain}}, \bibinfo {author} {\bibfnamefont
  {L.}~\bibnamefont {Chen}}, \bibinfo {author} {\bibfnamefont {G.~J.}\
  \bibnamefont {Snyder}},\ and\ \bibinfo {author} {\bibfnamefont
  {Y.}~\bibnamefont {Pei}},\ }\bibfield  {title} {\bibinfo {title}
  {Low-symmetry rhombohedral {GeTe} thermoelectrics},\ }\href
  {https://doi.org/10.1016/j.joule.2018.02.016} {\bibfield  {journal} {\bibinfo
   {journal} {Joule}\ }\textbf {\bibinfo {volume} {2}},\ \bibinfo {pages} {976}
  (\bibinfo {year} {2018}{\natexlab{b}})}\BibitemShut {NoStop}%
\bibitem [{\citenamefont {Liu}\ \emph {et~al.}(2020)\citenamefont {Liu},
  \citenamefont {Wang}, \citenamefont {Liu}, \citenamefont {Zhou},
  \citenamefont {Shao},\ and\ \citenamefont {Chen}}]{liu2020high}%
  \BibitemOpen
  \bibfield  {author} {\bibinfo {author} {\bibfnamefont {W.-D.}\ \bibnamefont
  {Liu}}, \bibinfo {author} {\bibfnamefont {D.-Z.}\ \bibnamefont {Wang}},
  \bibinfo {author} {\bibfnamefont {Q.}~\bibnamefont {Liu}}, \bibinfo {author}
  {\bibfnamefont {W.}~\bibnamefont {Zhou}}, \bibinfo {author} {\bibfnamefont
  {Z.}~\bibnamefont {Shao}},\ and\ \bibinfo {author} {\bibfnamefont {Z.-G.}\
  \bibnamefont {Chen}},\ }\bibfield  {title} {\bibinfo {title}
  {{High-Performance} {GeTe-Based} thermoelectrics: {From} materials to
  devices},\ }\href {https://doi.org/10.1002/aenm.202000367} {\bibfield
  {journal} {\bibinfo  {journal} {Adv. Energy Mater.}\ }\textbf {\bibinfo
  {volume} {10}},\ \bibinfo {pages} {2000367} (\bibinfo {year}
  {2020})}\BibitemShut {NoStop}%
\bibitem [{\citenamefont {Lee}\ \emph {et~al.}(2014)\citenamefont {Lee},
  \citenamefont {Esfarjani}, \citenamefont {Luo}, \citenamefont {Zhou},
  \citenamefont {Tian},\ and\ \citenamefont {Chen}}]{lee2014resonant}%
  \BibitemOpen
  \bibfield  {author} {\bibinfo {author} {\bibfnamefont {S.}~\bibnamefont
  {Lee}}, \bibinfo {author} {\bibfnamefont {K.}~\bibnamefont {Esfarjani}},
  \bibinfo {author} {\bibfnamefont {T.}~\bibnamefont {Luo}}, \bibinfo {author}
  {\bibfnamefont {J.}~\bibnamefont {Zhou}}, \bibinfo {author} {\bibfnamefont
  {Z.}~\bibnamefont {Tian}},\ and\ \bibinfo {author} {\bibfnamefont
  {G.}~\bibnamefont {Chen}},\ }\bibfield  {title} {\bibinfo {title} {Resonant
  bonding leads to low lattice thermal conductivity},\ }\href
  {https://doi.org/10.1038/ncomms4525} {\bibfield  {journal} {\bibinfo
  {journal} {Nat. Commun.}\ }\textbf {\bibinfo {volume} {5}},\ \bibinfo {pages}
  {3525} (\bibinfo {year} {2014})}\BibitemShut {NoStop}%
\bibitem [{\citenamefont {Yue}\ \emph {et~al.}(2023)\citenamefont {Yue},
  \citenamefont {Zhao}, \citenamefont {Ni}, \citenamefont {Meng},\ and\
  \citenamefont {Dai}}]{yue2023strong}%
  \BibitemOpen
  \bibfield  {author} {\bibinfo {author} {\bibfnamefont {T.}~\bibnamefont
  {Yue}}, \bibinfo {author} {\bibfnamefont {Y.}~\bibnamefont {Zhao}}, \bibinfo
  {author} {\bibfnamefont {J.}~\bibnamefont {Ni}}, \bibinfo {author}
  {\bibfnamefont {S.}~\bibnamefont {Meng}},\ and\ \bibinfo {author}
  {\bibfnamefont {Z.}~\bibnamefont {Dai}},\ }\bibfield  {title} {\bibinfo
  {title} {Strong quartic anharmonicity, ultralow thermal conductivity, high
  band degeneracy and good thermoelectric performance in {Na$_2$TlSb}},\ }\href
  {https://doi.org/10.1038/s41524-023-00970-4} {\bibfield  {journal} {\bibinfo
  {journal} {npj Comput. Mater.}\ }\textbf {\bibinfo {volume} {9}},\ \bibinfo
  {pages} {17} (\bibinfo {year} {2023})}\BibitemShut {NoStop}%
\bibitem [{\citenamefont {Zhang}\ \emph {et~al.}(2023)\citenamefont {Zhang},
  \citenamefont {Ishikawa}, \citenamefont {Koza}, \citenamefont {Nishibori},
  \citenamefont {Song}, \citenamefont {Baron},\ and\ \citenamefont
  {Iversen}}]{zhang2023dynamic}%
  \BibitemOpen
  \bibfield  {author} {\bibinfo {author} {\bibfnamefont {J.}~\bibnamefont
  {Zhang}}, \bibinfo {author} {\bibfnamefont {D.}~\bibnamefont {Ishikawa}},
  \bibinfo {author} {\bibfnamefont {M.~M.}\ \bibnamefont {Koza}}, \bibinfo
  {author} {\bibfnamefont {E.}~\bibnamefont {Nishibori}}, \bibinfo {author}
  {\bibfnamefont {L.}~\bibnamefont {Song}}, \bibinfo {author} {\bibfnamefont
  {A.~Q.~R.}\ \bibnamefont {Baron}},\ and\ \bibinfo {author} {\bibfnamefont
  {B.~B.}\ \bibnamefont {Iversen}},\ }\bibfield  {title} {\bibinfo {title}
  {Dynamic lone pair expression as chemical bonding origin of giant phonon
  anharmonicity in thermoelectric {InTe}},\ }\href
  {https://doi.org/10.1002/ange.202218458} {\bibfield  {journal} {\bibinfo
  {journal} {Angew. Chem.}\ }\textbf {\bibinfo {volume} {135}},\ \bibinfo
  {pages} {e202218458} (\bibinfo {year} {2023})}\BibitemShut {NoStop}%
\bibitem [{\citenamefont {Wan}\ \emph {et~al.}(2019)\citenamefont {Wan},
  \citenamefont {Ge},\ and\ \citenamefont {Liu}}]{wan2019strong}%
  \BibitemOpen
  \bibfield  {author} {\bibinfo {author} {\bibfnamefont {W.}~\bibnamefont
  {Wan}}, \bibinfo {author} {\bibfnamefont {Y.}~\bibnamefont {Ge}},\ and\
  \bibinfo {author} {\bibfnamefont {Y.}~\bibnamefont {Liu}},\ }\bibfield
  {title} {\bibinfo {title} {Strong phonon anharmonicity and low thermal
  conductivity of monolayer tin oxides driven by lone-pair electrons},\ }\href
  {https://doi.org/10.1063/1.5063560} {\bibfield  {journal} {\bibinfo
  {journal} {Appl. Phys. Lett.}\ }\textbf {\bibinfo {volume} {114}},\ \bibinfo
  {pages} {031901} (\bibinfo {year} {2019})}\BibitemShut {NoStop}%
\bibitem [{\citenamefont {Kim}\ \emph {et~al.}(2020)\citenamefont {Kim},
  \citenamefont {Lee}, \citenamefont {Jung}, \citenamefont {Shin},\ and\
  \citenamefont {Park}}]{kim2020high}%
  \BibitemOpen
  \bibfield  {author} {\bibinfo {author} {\bibfnamefont {J.~Y.}\ \bibnamefont
  {Kim}}, \bibinfo {author} {\bibfnamefont {J.-W.}\ \bibnamefont {Lee}},
  \bibinfo {author} {\bibfnamefont {H.~S.}\ \bibnamefont {Jung}}, \bibinfo
  {author} {\bibfnamefont {H.}~\bibnamefont {Shin}},\ and\ \bibinfo {author}
  {\bibfnamefont {N.-G.}\ \bibnamefont {Park}},\ }\bibfield  {title} {\bibinfo
  {title} {High-efficiency perovskite solar cells},\ }\href
  {https://doi.org/10.1021/acs.chemrev.0c00107} {\bibfield  {journal} {\bibinfo
   {journal} {Chem. Rev.}\ }\textbf {\bibinfo {volume} {120}},\ \bibinfo
  {pages} {7867} (\bibinfo {year} {2020})}\BibitemShut {NoStop}%
\bibitem [{\citenamefont {Ma}\ \emph {et~al.}(2019)\citenamefont {Ma},
  \citenamefont {Li}, \citenamefont {Ma}, \citenamefont {Wang}, \citenamefont
  {Rouse}, \citenamefont {Zhang}, \citenamefont {Slebodnick}, \citenamefont
  {Alatas}, \citenamefont {Baker}, \citenamefont {Urban},\ and\ \citenamefont
  {Tian}}]{ma2019supercompliant}%
  \BibitemOpen
  \bibfield  {author} {\bibinfo {author} {\bibfnamefont {H.}~\bibnamefont
  {Ma}}, \bibinfo {author} {\bibfnamefont {C.}~\bibnamefont {Li}}, \bibinfo
  {author} {\bibfnamefont {Y.}~\bibnamefont {Ma}}, \bibinfo {author}
  {\bibfnamefont {H.}~\bibnamefont {Wang}}, \bibinfo {author} {\bibfnamefont
  {Z.~W.}\ \bibnamefont {Rouse}}, \bibinfo {author} {\bibfnamefont
  {Z.}~\bibnamefont {Zhang}}, \bibinfo {author} {\bibfnamefont
  {C.}~\bibnamefont {Slebodnick}}, \bibinfo {author} {\bibfnamefont
  {A.}~\bibnamefont {Alatas}}, \bibinfo {author} {\bibfnamefont {S.~P.}\
  \bibnamefont {Baker}}, \bibinfo {author} {\bibfnamefont {J.~J.}\ \bibnamefont
  {Urban}},\ and\ \bibinfo {author} {\bibfnamefont {Z.}~\bibnamefont {Tian}},\
  }\bibfield  {title} {\bibinfo {title} {Supercompliant and soft
  {(CH$_3$NH$_3$)$_3$Bi$_2$I$_9$} crystal with ultralow thermal conductivity},\
  }\href {https://doi.org/10.1103/physrevlett.123.155901} {\bibfield  {journal}
  {\bibinfo  {journal} {Phys. Rev. Lett.}\ }\textbf {\bibinfo {volume} {123}},\
  \bibinfo {pages} {155901} (\bibinfo {year} {2019})}\BibitemShut {NoStop}%
\bibitem [{\citenamefont {Fabini}\ \emph {et~al.}(2020)\citenamefont {Fabini},
  \citenamefont {Seshadri},\ and\ \citenamefont
  {Kanatzidis}}]{fabini2020underappreciated}%
  \BibitemOpen
  \bibfield  {author} {\bibinfo {author} {\bibfnamefont {D.~H.}\ \bibnamefont
  {Fabini}}, \bibinfo {author} {\bibfnamefont {R.}~\bibnamefont {Seshadri}},\
  and\ \bibinfo {author} {\bibfnamefont {M.~G.}\ \bibnamefont {Kanatzidis}},\
  }\bibfield  {title} {\bibinfo {title} {The underappreciated lone pair in
  halide perovskites underpins their unusual properties},\ }\href
  {https://doi.org/10.1557/mrs.2020.142} {\bibfield  {journal} {\bibinfo
  {journal} {MRS Bull.}\ }\textbf {\bibinfo {volume} {45}},\ \bibinfo {pages}
  {467} (\bibinfo {year} {2020})}\BibitemShut {NoStop}%
\bibitem [{\citenamefont {Kim}\ \emph {et~al.}(2023)\citenamefont {Kim},
  \citenamefont {Park}, \citenamefont {Iyer}, \citenamefont {Shaheen},
  \citenamefont {Choudhry}, \citenamefont {Jiang}, \citenamefont {Eichman},
  \citenamefont {Gnabasik}, \citenamefont {Kelley}, \citenamefont {Lawrie},
  \citenamefont {Zhu},\ and\ \citenamefont {Liao}}]{kim2023mapping}%
  \BibitemOpen
  \bibfield  {author} {\bibinfo {author} {\bibfnamefont {T.}~\bibnamefont
  {Kim}}, \bibinfo {author} {\bibfnamefont {S.}~\bibnamefont {Park}}, \bibinfo
  {author} {\bibfnamefont {V.}~\bibnamefont {Iyer}}, \bibinfo {author}
  {\bibfnamefont {B.}~\bibnamefont {Shaheen}}, \bibinfo {author} {\bibfnamefont
  {U.}~\bibnamefont {Choudhry}}, \bibinfo {author} {\bibfnamefont
  {Q.}~\bibnamefont {Jiang}}, \bibinfo {author} {\bibfnamefont
  {G.}~\bibnamefont {Eichman}}, \bibinfo {author} {\bibfnamefont
  {R.}~\bibnamefont {Gnabasik}}, \bibinfo {author} {\bibfnamefont
  {K.}~\bibnamefont {Kelley}}, \bibinfo {author} {\bibfnamefont
  {B.}~\bibnamefont {Lawrie}}, \bibinfo {author} {\bibfnamefont
  {K.}~\bibnamefont {Zhu}},\ and\ \bibinfo {author} {\bibfnamefont
  {B.}~\bibnamefont {Liao}},\ }\bibfield  {title} {\bibinfo {title} {Mapping
  the pathways of photo-induced ion migration in organic-inorganic hybrid
  halide perovskites},\ }\href {https://doi.org/10.1038/s41467-023-37486-w}
  {\bibfield  {journal} {\bibinfo  {journal} {Nat. Commun.}\ }\textbf {\bibinfo
  {volume} {14}},\ \bibinfo {pages} {1846} (\bibinfo {year}
  {2023})}\BibitemShut {NoStop}%
\bibitem [{\citenamefont {Wang}\ \emph {et~al.}(2020)\citenamefont {Wang},
  \citenamefont {Xiong}, \citenamefont {Wang}, \citenamefont {Qiu},
  \citenamefont {Song}, \citenamefont {Zhao}, \citenamefont {Yang},
  \citenamefont {Xiao}, \citenamefont {Shi},\ and\ \citenamefont
  {Chen}}]{wang2020cu3erte3}%
  \BibitemOpen
  \bibfield  {author} {\bibinfo {author} {\bibfnamefont {T.}~\bibnamefont
  {Wang}}, \bibinfo {author} {\bibfnamefont {Y.}~\bibnamefont {Xiong}},
  \bibinfo {author} {\bibfnamefont {Y.}~\bibnamefont {Wang}}, \bibinfo {author}
  {\bibfnamefont {P.}~\bibnamefont {Qiu}}, \bibinfo {author} {\bibfnamefont
  {Q.}~\bibnamefont {Song}}, \bibinfo {author} {\bibfnamefont {K.}~\bibnamefont
  {Zhao}}, \bibinfo {author} {\bibfnamefont {J.}~\bibnamefont {Yang}}, \bibinfo
  {author} {\bibfnamefont {J.}~\bibnamefont {Xiao}}, \bibinfo {author}
  {\bibfnamefont {X.}~\bibnamefont {Shi}},\ and\ \bibinfo {author}
  {\bibfnamefont {L.}~\bibnamefont {Chen}},\ }\bibfield  {title} {\bibinfo
  {title} {{Cu$_3$ErTe$_3$:} {A} new promising thermoelectric material
  predicated by high-throughput screening},\ }\href
  {https://doi.org/10.1016/j.mtphys.2020.100180} {\bibfield  {journal}
  {\bibinfo  {journal} {Mater. Today Phys.}\ }\textbf {\bibinfo {volume}
  {12}},\ \bibinfo {pages} {100180} (\bibinfo {year} {2020})}\BibitemShut
  {NoStop}%
\bibitem [{\citenamefont {Li}\ \emph {et~al.}(2022{\natexlab{a}})\citenamefont
  {Li}, \citenamefont {Hu},\ and\ \citenamefont {Yang}}]{li2022high}%
  \BibitemOpen
  \bibfield  {author} {\bibinfo {author} {\bibfnamefont {J.}~\bibnamefont
  {Li}}, \bibinfo {author} {\bibfnamefont {W.}~\bibnamefont {Hu}},\ and\
  \bibinfo {author} {\bibfnamefont {J.}~\bibnamefont {Yang}},\ }\bibfield
  {title} {\bibinfo {title} {High-throughput screening of rattling-induced
  ultralow lattice thermal conductivity in semiconductors},\ }\href
  {https://doi.org/10.1021/jacs.1c11887} {\bibfield  {journal} {\bibinfo
  {journal} {JACS}\ }\textbf {\bibinfo {volume} {144}},\ \bibinfo {pages}
  {4448} (\bibinfo {year} {2022}{\natexlab{a}})}\BibitemShut {NoStop}%
\bibitem [{\citenamefont {Carrete}\ \emph {et~al.}(2014)\citenamefont
  {Carrete}, \citenamefont {Li}, \citenamefont {Mingo}, \citenamefont {Wang},\
  and\ \citenamefont {Curtarolo}}]{carrete2014finding}%
  \BibitemOpen
  \bibfield  {author} {\bibinfo {author} {\bibfnamefont {J.}~\bibnamefont
  {Carrete}}, \bibinfo {author} {\bibfnamefont {W.}~\bibnamefont {Li}},
  \bibinfo {author} {\bibfnamefont {N.}~\bibnamefont {Mingo}}, \bibinfo
  {author} {\bibfnamefont {S.}~\bibnamefont {Wang}},\ and\ \bibinfo {author}
  {\bibfnamefont {S.}~\bibnamefont {Curtarolo}},\ }\bibfield  {title} {\bibinfo
  {title} {Finding unprecedentedly low-thermal-conductivity half-heusler
  semiconductors via high-throughput materials modeling},\ }\href
  {https://doi.org/10.1103/physrevx.4.011019} {\bibfield  {journal} {\bibinfo
  {journal} {Phys. Rev. X}\ }\textbf {\bibinfo {volume} {4}},\ \bibinfo {pages}
  {011019} (\bibinfo {year} {2014})}\BibitemShut {NoStop}%
\bibitem [{\citenamefont {Rodriguez}\ \emph {et~al.}(2023)\citenamefont
  {Rodriguez}, \citenamefont {Lin}, \citenamefont {Yang}, \citenamefont
  {Al-Fahdi}, \citenamefont {Shen}, \citenamefont {Choudhary}, \citenamefont
  {Zhao}, \citenamefont {Hu}, \citenamefont {Cao}, \citenamefont {Zhang},\ and\
  \citenamefont {Hu}}]{rodriguez2023million}%
  \BibitemOpen
  \bibfield  {author} {\bibinfo {author} {\bibfnamefont {A.}~\bibnamefont
  {Rodriguez}}, \bibinfo {author} {\bibfnamefont {C.}~\bibnamefont {Lin}},
  \bibinfo {author} {\bibfnamefont {H.}~\bibnamefont {Yang}}, \bibinfo {author}
  {\bibfnamefont {M.}~\bibnamefont {Al-Fahdi}}, \bibinfo {author}
  {\bibfnamefont {C.}~\bibnamefont {Shen}}, \bibinfo {author} {\bibfnamefont
  {K.}~\bibnamefont {Choudhary}}, \bibinfo {author} {\bibfnamefont
  {Y.}~\bibnamefont {Zhao}}, \bibinfo {author} {\bibfnamefont {J.}~\bibnamefont
  {Hu}}, \bibinfo {author} {\bibfnamefont {B.}~\bibnamefont {Cao}}, \bibinfo
  {author} {\bibfnamefont {H.}~\bibnamefont {Zhang}},\ and\ \bibinfo {author}
  {\bibfnamefont {M.}~\bibnamefont {Hu}},\ }\bibfield  {title} {\bibinfo
  {title} {Million-scale data integrated deep neural network for phonon
  properties of heuslers spanning the periodic table},\ }\href
  {https://doi.org/10.1038/s41524-023-00974-0} {\bibfield  {journal} {\bibinfo
  {journal} {npj Comput. Mater.}\ }\textbf {\bibinfo {volume} {9}},\ \bibinfo
  {pages} {20} (\bibinfo {year} {2023})}\BibitemShut {NoStop}%
\bibitem [{\citenamefont {Yuan}\ \emph {et~al.}(2022)\citenamefont {Yuan},
  \citenamefont {Zhang}, \citenamefont {Chang}, \citenamefont {Tang},\ and\
  \citenamefont {Hu}}]{yuan2022antibonding}%
  \BibitemOpen
  \bibfield  {author} {\bibinfo {author} {\bibfnamefont {K.}~\bibnamefont
  {Yuan}}, \bibinfo {author} {\bibfnamefont {X.}~\bibnamefont {Zhang}},
  \bibinfo {author} {\bibfnamefont {Z.}~\bibnamefont {Chang}}, \bibinfo
  {author} {\bibfnamefont {D.}~\bibnamefont {Tang}},\ and\ \bibinfo {author}
  {\bibfnamefont {M.}~\bibnamefont {Hu}},\ }\bibfield  {title} {\bibinfo
  {title} {Antibonding induced anharmonicity leading to ultralow lattice
  thermal conductivity and extraordinary thermoelectric performance in
  {CsK$_2$X} ({X} = {Sb}, {Bi})},\ }\href {https://doi.org/10.1039/d2tc03356a}
  {\bibfield  {journal} {\bibinfo  {journal} {J. Mater. Chem. C}\ }\textbf
  {\bibinfo {volume} {10}},\ \bibinfo {pages} {15822} (\bibinfo {year}
  {2022})}\BibitemShut {NoStop}%
\bibitem [{\citenamefont {Das}\ \emph {et~al.}(2023)\citenamefont {Das},
  \citenamefont {Pal}, \citenamefont {Acharyya}, \citenamefont {Das},
  \citenamefont {Maji},\ and\ \citenamefont {Biswas}}]{das2023strong}%
  \BibitemOpen
  \bibfield  {author} {\bibinfo {author} {\bibfnamefont {A.}~\bibnamefont
  {Das}}, \bibinfo {author} {\bibfnamefont {K.}~\bibnamefont {Pal}}, \bibinfo
  {author} {\bibfnamefont {P.}~\bibnamefont {Acharyya}}, \bibinfo {author}
  {\bibfnamefont {S.}~\bibnamefont {Das}}, \bibinfo {author} {\bibfnamefont
  {K.}~\bibnamefont {Maji}},\ and\ \bibinfo {author} {\bibfnamefont
  {K.}~\bibnamefont {Biswas}},\ }\bibfield  {title} {\bibinfo {title} {Strong
  antibonding {I (p){\textendash}Cu (d)} states lead to intrinsically low
  thermal conductivity in {CuBiI$_4$}},\ }\href
  {https://doi.org/10.1021/jacs.2c11908} {\bibfield  {journal} {\bibinfo
  {journal} {JACS}\ }\textbf {\bibinfo {volume} {145}},\ \bibinfo {pages}
  {1349} (\bibinfo {year} {2023})}\BibitemShut {NoStop}%
\bibitem [{\citenamefont {Zhang}\ \emph {et~al.}(2022)\citenamefont {Zhang},
  \citenamefont {Jiang}, \citenamefont {Xia}, \citenamefont {Gao},\ and\
  \citenamefont {Huang}}]{zhang2022antibonding}%
  \BibitemOpen
  \bibfield  {author} {\bibinfo {author} {\bibfnamefont {J.}~\bibnamefont
  {Zhang}}, \bibinfo {author} {\bibfnamefont {H.}~\bibnamefont {Jiang}},
  \bibinfo {author} {\bibfnamefont {X.}~\bibnamefont {Xia}}, \bibinfo {author}
  {\bibfnamefont {Y.}~\bibnamefont {Gao}},\ and\ \bibinfo {author}
  {\bibfnamefont {Z.}~\bibnamefont {Huang}},\ }\bibfield  {title} {\bibinfo
  {title} {Antibonding p-d and s-p hybridization induce the optimization of
  thermal and thermoelectric performance of {MGeTe$_3$} ({M} = {In} and
  {Sb})},\ }\href {https://doi.org/10.1021/acsaem.2c03138} {\bibfield
  {journal} {\bibinfo  {journal} {ACS Appl. Energy Mater.}\ }\textbf {\bibinfo
  {volume} {5}},\ \bibinfo {pages} {15566} (\bibinfo {year}
  {2022})}\BibitemShut {NoStop}%
\bibitem [{\citenamefont {Dronskowski}\ and\ \citenamefont
  {Bloechl}(1993)}]{dronskowski1993crystal}%
  \BibitemOpen
  \bibfield  {author} {\bibinfo {author} {\bibfnamefont {R.}~\bibnamefont
  {Dronskowski}}\ and\ \bibinfo {author} {\bibfnamefont {P.~E.}\ \bibnamefont
  {Bloechl}},\ }\bibfield  {title} {\bibinfo {title} {Crystal orbital
  {Hamilton} populations {(COHP):} {Energy-resolved} visualization of chemical
  bonding in solids based on density-functional calculations},\ }\href
  {https://doi.org/10.1021/j100135a014} {\bibfield  {journal} {\bibinfo
  {journal} {The Journal of Physical Chemistry}\ }\textbf {\bibinfo {volume}
  {97}},\ \bibinfo {pages} {8617} (\bibinfo {year} {1993})}\BibitemShut
  {NoStop}%
\bibitem [{\citenamefont {Deringer}\ \emph {et~al.}(2011)\citenamefont
  {Deringer}, \citenamefont {Tchougr\'eeff},\ and\ \citenamefont
  {Dronskowski}}]{deringer2011crystal}%
  \BibitemOpen
  \bibfield  {author} {\bibinfo {author} {\bibfnamefont {V.~L.}\ \bibnamefont
  {Deringer}}, \bibinfo {author} {\bibfnamefont {A.~L.}\ \bibnamefont
  {Tchougr\'eeff}},\ and\ \bibinfo {author} {\bibfnamefont {R.}~\bibnamefont
  {Dronskowski}},\ }\bibfield  {title} {\bibinfo {title} {Crystal orbital
  {Hamilton} population {(COHP)} analysis as projected from plane-wave basis
  sets},\ }\href {https://doi.org/10.1021/jp202489s} {\bibfield  {journal}
  {\bibinfo  {journal} {J. Phys. Chem. A}\ }\textbf {\bibinfo {volume} {115}},\
  \bibinfo {pages} {5461} (\bibinfo {year} {2011})}\BibitemShut {NoStop}%
\bibitem [{\citenamefont {Jain}\ \emph {et~al.}(2013)\citenamefont {Jain},
  \citenamefont {Ong}, \citenamefont {Hautier}, \citenamefont {Chen},
  \citenamefont {Richards}, \citenamefont {Dacek}, \citenamefont {Cholia},
  \citenamefont {Gunter}, \citenamefont {Skinner}, \citenamefont {Ceder},\ and\
  \citenamefont {Persson}}]{jain2013commentary}%
  \BibitemOpen
  \bibfield  {author} {\bibinfo {author} {\bibfnamefont {A.}~\bibnamefont
  {Jain}}, \bibinfo {author} {\bibfnamefont {S.~P.}\ \bibnamefont {Ong}},
  \bibinfo {author} {\bibfnamefont {G.}~\bibnamefont {Hautier}}, \bibinfo
  {author} {\bibfnamefont {W.}~\bibnamefont {Chen}}, \bibinfo {author}
  {\bibfnamefont {W.~D.}\ \bibnamefont {Richards}}, \bibinfo {author}
  {\bibfnamefont {S.}~\bibnamefont {Dacek}}, \bibinfo {author} {\bibfnamefont
  {S.}~\bibnamefont {Cholia}}, \bibinfo {author} {\bibfnamefont
  {D.}~\bibnamefont {Gunter}}, \bibinfo {author} {\bibfnamefont
  {D.}~\bibnamefont {Skinner}}, \bibinfo {author} {\bibfnamefont
  {G.}~\bibnamefont {Ceder}},\ and\ \bibinfo {author} {\bibfnamefont {K.~A.}\
  \bibnamefont {Persson}},\ }\bibfield  {title} {\bibinfo {title} {Commentary:
  {The} materials project: {A} materials genome approach to accelerating
  materials innovation},\ }\href {https://doi.org/10.1063/1.4812323} {\bibfield
   {journal} {\bibinfo  {journal} {APL Mater.}\ }\textbf {\bibinfo {volume}
  {1}},\ \bibinfo {pages} {011002} (\bibinfo {year} {2013})}\BibitemShut
  {NoStop}%
\bibitem [{\citenamefont {Kresse}\ and\ \citenamefont
  {Furthm\"uller}(1996{\natexlab{a}})}]{kresse1996vasp1}%
  \BibitemOpen
  \bibfield  {author} {\bibinfo {author} {\bibfnamefont {G.}~\bibnamefont
  {Kresse}}\ and\ \bibinfo {author} {\bibfnamefont {J.}~\bibnamefont
  {Furthm\"uller}},\ }\bibfield  {title} {\bibinfo {title} {Efficient iterative
  schemes for \textit{ab initio} total-energy calculations using a plane-wave
  basis set},\ }\href {https://doi.org/10.1103/physrevb.54.11169} {\bibfield
  {journal} {\bibinfo  {journal} {Phys. Rev. B}\ }\textbf {\bibinfo {volume}
  {54}},\ \bibinfo {pages} {11169} (\bibinfo {year}
  {1996}{\natexlab{a}})}\BibitemShut {NoStop}%
\bibitem [{\citenamefont {Kresse}\ and\ \citenamefont
  {Furthm\"uller}(1996{\natexlab{b}})}]{kresse1996vasp2}%
  \BibitemOpen
  \bibfield  {author} {\bibinfo {author} {\bibfnamefont {G.}~\bibnamefont
  {Kresse}}\ and\ \bibinfo {author} {\bibfnamefont {J.}~\bibnamefont
  {Furthm\"uller}},\ }\bibfield  {title} {\bibinfo {title} {Efficiency of
  ab-initio total energy calculations for metals and semiconductors using a
  plane-wave basis set},\ }\href {https://doi.org/10.1016/0927-0256(96)00008-0}
  {\bibfield  {journal} {\bibinfo  {journal} {Comput. Mater. Sci.}\ }\textbf
  {\bibinfo {volume} {6}},\ \bibinfo {pages} {15} (\bibinfo {year}
  {1996}{\natexlab{b}})}\BibitemShut {NoStop}%
\bibitem [{\citenamefont {Bl\"ochl}(1994)}]{blochl1994projector}%
  \BibitemOpen
  \bibfield  {author} {\bibinfo {author} {\bibfnamefont {P.~E.}\ \bibnamefont
  {Bl\"ochl}},\ }\bibfield  {title} {\bibinfo {title} {Projector augmented-wave
  method},\ }\href {https://doi.org/10.1103/physrevb.50.17953} {\bibfield
  {journal} {\bibinfo  {journal} {Phys. Rev. B}\ }\textbf {\bibinfo {volume}
  {50}},\ \bibinfo {pages} {17953} (\bibinfo {year} {1994})}\BibitemShut
  {NoStop}%
\bibitem [{\citenamefont {Perdew}\ \emph {et~al.}(1996)\citenamefont {Perdew},
  \citenamefont {Burke},\ and\ \citenamefont
  {Ernzerhof}}]{perdew1996generalized}%
  \BibitemOpen
  \bibfield  {author} {\bibinfo {author} {\bibfnamefont {J.~P.}\ \bibnamefont
  {Perdew}}, \bibinfo {author} {\bibfnamefont {K.}~\bibnamefont {Burke}},\ and\
  \bibinfo {author} {\bibfnamefont {M.}~\bibnamefont {Ernzerhof}},\ }\bibfield
  {title} {\bibinfo {title} {Generalized gradient approximation made simple},\
  }\href {https://doi.org/10.1103/physrevlett.77.3865} {\bibfield  {journal}
  {\bibinfo  {journal} {Phys. Rev. Lett.}\ }\textbf {\bibinfo {volume} {77}},\
  \bibinfo {pages} {3865} (\bibinfo {year} {1996})}\BibitemShut {NoStop}%
\bibitem [{\citenamefont {Broido}\ \emph {et~al.}(2007)\citenamefont {Broido},
  \citenamefont {Malorny}, \citenamefont {Birner}, \citenamefont {Mingo},\ and\
  \citenamefont {Stewart}}]{broido2007intrinsic}%
  \BibitemOpen
  \bibfield  {author} {\bibinfo {author} {\bibfnamefont {D.~A.}\ \bibnamefont
  {Broido}}, \bibinfo {author} {\bibfnamefont {M.}~\bibnamefont {Malorny}},
  \bibinfo {author} {\bibfnamefont {G.}~\bibnamefont {Birner}}, \bibinfo
  {author} {\bibfnamefont {N.}~\bibnamefont {Mingo}},\ and\ \bibinfo {author}
  {\bibfnamefont {D.~A.}\ \bibnamefont {Stewart}},\ }\bibfield  {title}
  {\bibinfo {title} {Intrinsic lattice thermal conductivity of semiconductors
  from first principles},\ }\href {https://doi.org/10.1063/1.2822891}
  {\bibfield  {journal} {\bibinfo  {journal} {Appl. Phys. Lett.}\ }\textbf
  {\bibinfo {volume} {91}},\ \bibinfo {pages} {231922} (\bibinfo {year}
  {2007})}\BibitemShut {NoStop}%
\bibitem [{\citenamefont {Togo}\ and\ \citenamefont
  {Tanaka}(2015)}]{togo2015first}%
  \BibitemOpen
  \bibfield  {author} {\bibinfo {author} {\bibfnamefont {A.}~\bibnamefont
  {Togo}}\ and\ \bibinfo {author} {\bibfnamefont {I.}~\bibnamefont {Tanaka}},\
  }\bibfield  {title} {\bibinfo {title} {First principles phonon calculations
  in materials science},\ }\href
  {https://doi.org/10.1016/j.scriptamat.2015.07.021} {\bibfield  {journal}
  {\bibinfo  {journal} {Scripta Mater.}\ }\textbf {\bibinfo {volume} {108}},\
  \bibinfo {pages} {1} (\bibinfo {year} {2015})}\BibitemShut {NoStop}%
\bibitem [{\citenamefont {Li}\ \emph {et~al.}(2014)\citenamefont {Li},
  \citenamefont {Carrete}, \citenamefont {A.~Katcho},\ and\ \citenamefont
  {Mingo}}]{li2014shengbte}%
  \BibitemOpen
  \bibfield  {author} {\bibinfo {author} {\bibfnamefont {W.}~\bibnamefont
  {Li}}, \bibinfo {author} {\bibfnamefont {J.}~\bibnamefont {Carrete}},
  \bibinfo {author} {\bibfnamefont {N.}~\bibnamefont {A.~Katcho}},\ and\
  \bibinfo {author} {\bibfnamefont {N.}~\bibnamefont {Mingo}},\ }\bibfield
  {title} {\bibinfo {title} {{ShengBTE:} {A} solver of the {Boltzmann}
  transport equation for phonons},\ }\href
  {https://doi.org/10.1016/j.cpc.2014.02.015} {\bibfield  {journal} {\bibinfo
  {journal} {Comput. Phys. Commun.}\ }\textbf {\bibinfo {volume} {185}},\
  \bibinfo {pages} {1747} (\bibinfo {year} {2014})}\BibitemShut {NoStop}%
\bibitem [{\citenamefont {Pople}\ and\ \citenamefont
  {Beveridge}(1970)}]{pople1970molecular}%
  \BibitemOpen
  \bibfield  {author} {\bibinfo {author} {\bibfnamefont {J.~A.}\ \bibnamefont
  {Pople}}\ and\ \bibinfo {author} {\bibfnamefont {D.~L.}\ \bibnamefont
  {Beveridge}},\ }\bibfield  {title} {\bibinfo {title} {Molecular orbital
  theory},\ }\href
  {http://www.sciencemadness.org/library/books/approximate_mo_theory.pdf}
  {\bibfield  {journal} {\bibinfo  {journal} {C0., NY}\ } (\bibinfo {year}
  {1970})}\BibitemShut {NoStop}%
\bibitem [{\citenamefont {Hehre}(1976)}]{hehre1976ab}%
  \BibitemOpen
  \bibfield  {author} {\bibinfo {author} {\bibfnamefont {W.~J.}\ \bibnamefont
  {Hehre}},\ }\bibfield  {title} {\bibinfo {title} {Ab initio molecular orbital
  theory},\ }\href {https://doi.org/10.1021/ar50107a003} {\bibfield  {journal}
  {\bibinfo  {journal} {Accounts Chem. Res.}\ }\textbf {\bibinfo {volume}
  {9}},\ \bibinfo {pages} {399} (\bibinfo {year} {1976})}\BibitemShut {NoStop}%
\bibitem [{\citenamefont {Heremans}\ \emph {et~al.}(2008)\citenamefont
  {Heremans}, \citenamefont {Jovovic}, \citenamefont {Toberer}, \citenamefont
  {Saramat}, \citenamefont {Kurosaki}, \citenamefont {Charoenphakdee},
  \citenamefont {Yamanaka},\ and\ \citenamefont
  {Snyder}}]{heremans2008enhancement}%
  \BibitemOpen
  \bibfield  {author} {\bibinfo {author} {\bibfnamefont {J.~P.}\ \bibnamefont
  {Heremans}}, \bibinfo {author} {\bibfnamefont {V.}~\bibnamefont {Jovovic}},
  \bibinfo {author} {\bibfnamefont {E.~S.}\ \bibnamefont {Toberer}}, \bibinfo
  {author} {\bibfnamefont {A.}~\bibnamefont {Saramat}}, \bibinfo {author}
  {\bibfnamefont {K.}~\bibnamefont {Kurosaki}}, \bibinfo {author}
  {\bibfnamefont {A.}~\bibnamefont {Charoenphakdee}}, \bibinfo {author}
  {\bibfnamefont {S.}~\bibnamefont {Yamanaka}},\ and\ \bibinfo {author}
  {\bibfnamefont {G.~J.}\ \bibnamefont {Snyder}},\ }\bibfield  {title}
  {\bibinfo {title} {Enhancement of thermoelectric efficiency in {PbTe} by
  distortion of the electronic density of states},\ }\href
  {https://doi.org/10.1126/science.1159725} {\bibfield  {journal} {\bibinfo
  {journal} {Science}\ }\textbf {\bibinfo {volume} {321}},\ \bibinfo {pages}
  {554} (\bibinfo {year} {2008})}\BibitemShut {NoStop}%
\bibitem [{\citenamefont {Akhmedova}\ and\ \citenamefont
  {Abdinov}(2009)}]{akhmedova2009effect}%
  \BibitemOpen
  \bibfield  {author} {\bibinfo {author} {\bibfnamefont {G.~A.}\ \bibnamefont
  {Akhmedova}}\ and\ \bibinfo {author} {\bibfnamefont {D.~S.}\ \bibnamefont
  {Abdinov}},\ }\bibfield  {title} {\bibinfo {title} {Effect of thallium doping
  on the thermal conductivity of {PbTe} single crystals},\ }\href
  {https://doi.org/10.1134/s0020168509080056} {\bibfield  {journal} {\bibinfo
  {journal} {Inorg. Mater.}\ }\textbf {\bibinfo {volume} {45}},\ \bibinfo
  {pages} {854} (\bibinfo {year} {2009})}\BibitemShut {NoStop}%
\bibitem [{\citenamefont {Waghmare}\ \emph {et~al.}(2003)\citenamefont
  {Waghmare}, \citenamefont {Spaldin}, \citenamefont {Kandpal},\ and\
  \citenamefont {Seshadri}}]{waghmare2003first}%
  \BibitemOpen
  \bibfield  {author} {\bibinfo {author} {\bibfnamefont {U.~V.}\ \bibnamefont
  {Waghmare}}, \bibinfo {author} {\bibfnamefont {N.~A.}\ \bibnamefont
  {Spaldin}}, \bibinfo {author} {\bibfnamefont {H.~C.}\ \bibnamefont
  {Kandpal}},\ and\ \bibinfo {author} {\bibfnamefont {R.}~\bibnamefont
  {Seshadri}},\ }\bibfield  {title} {\bibinfo {title} {First-principles
  indicators of metallicity and cation off-centricity in the {IV}-{VI} rocksalt
  chalcogenides of divalent {Ge}, {Sn}, and {Pb}},\ }\href
  {https://doi.org/10.1103/physrevb.67.125111} {\bibfield  {journal} {\bibinfo
  {journal} {Phys. Rev. B}\ }\textbf {\bibinfo {volume} {67}},\ \bibinfo
  {pages} {125111} (\bibinfo {year} {2003})}\BibitemShut {NoStop}%
\bibitem [{SM()}]{SM}%
  \BibitemOpen
  \href@noop {} {}\bibinfo {note} {See Supplemental Material at [URL will be
  inserted by publisher] for details about removing electrons in silicon,
  scattering properties of PbTe, thermal properties of XS family, screened list
  of anti-bonding binary semiconductors.}\BibitemShut {Stop}%
\bibitem [{\citenamefont {Jung}\ \emph {et~al.}(2019)\citenamefont {Jung},
  \citenamefont {Jeon}, \citenamefont {Park}, \citenamefont {Moon},
  \citenamefont {Shin}, \citenamefont {Yang}, \citenamefont {Noh},\ and\
  \citenamefont {Seo}}]{jung2019efficient}%
  \BibitemOpen
  \bibfield  {author} {\bibinfo {author} {\bibfnamefont {E.~H.}\ \bibnamefont
  {Jung}}, \bibinfo {author} {\bibfnamefont {N.~J.}\ \bibnamefont {Jeon}},
  \bibinfo {author} {\bibfnamefont {E.~Y.}\ \bibnamefont {Park}}, \bibinfo
  {author} {\bibfnamefont {C.~S.}\ \bibnamefont {Moon}}, \bibinfo {author}
  {\bibfnamefont {T.~J.}\ \bibnamefont {Shin}}, \bibinfo {author}
  {\bibfnamefont {T.-Y.}\ \bibnamefont {Yang}}, \bibinfo {author}
  {\bibfnamefont {J.~H.}\ \bibnamefont {Noh}},\ and\ \bibinfo {author}
  {\bibfnamefont {J.}~\bibnamefont {Seo}},\ }\bibfield  {title} {\bibinfo
  {title} {Efficient, stable and scalable perovskite solar cells using
  poly(3-hexylthiophene)},\ }\href {https://doi.org/10.1038/s41586-019-1036-3}
  {\bibfield  {journal} {\bibinfo  {journal} {Nature}\ }\textbf {\bibinfo
  {volume} {567}},\ \bibinfo {pages} {511} (\bibinfo {year}
  {2019})}\BibitemShut {NoStop}%
\bibitem [{\citenamefont {Li}\ \emph {et~al.}(2022{\natexlab{b}})\citenamefont
  {Li}, \citenamefont {Li}, \citenamefont {Wu}, \citenamefont {Sheppard},
  \citenamefont {Zhang}, \citenamefont {Gao}, \citenamefont {Long},\ and\
  \citenamefont {Zhu}}]{li2022organometallic}%
  \BibitemOpen
  \bibfield  {author} {\bibinfo {author} {\bibfnamefont {Z.}~\bibnamefont
  {Li}}, \bibinfo {author} {\bibfnamefont {B.}~\bibnamefont {Li}}, \bibinfo
  {author} {\bibfnamefont {X.}~\bibnamefont {Wu}}, \bibinfo {author}
  {\bibfnamefont {S.~A.}\ \bibnamefont {Sheppard}}, \bibinfo {author}
  {\bibfnamefont {S.}~\bibnamefont {Zhang}}, \bibinfo {author} {\bibfnamefont
  {D.}~\bibnamefont {Gao}}, \bibinfo {author} {\bibfnamefont {N.~J.}\
  \bibnamefont {Long}},\ and\ \bibinfo {author} {\bibfnamefont
  {Z.}~\bibnamefont {Zhu}},\ }\bibfield  {title} {\bibinfo {title}
  {Organometallic-functionalized interfaces for highly efficient inverted
  perovskite solar cells},\ }\href {https://doi.org/10.1126/science.abm8566}
  {\bibfield  {journal} {\bibinfo  {journal} {Science}\ }\textbf {\bibinfo
  {volume} {376}},\ \bibinfo {pages} {416} (\bibinfo {year}
  {2022}{\natexlab{b}})}\BibitemShut {NoStop}%
\bibitem [{\citenamefont {Haque}\ \emph {et~al.}(2020)\citenamefont {Haque},
  \citenamefont {Kee}, \citenamefont {Villalva}, \citenamefont {Ong},\ and\
  \citenamefont {Baran}}]{haque2020halide}%
  \BibitemOpen
  \bibfield  {author} {\bibinfo {author} {\bibfnamefont {M.~A.}\ \bibnamefont
  {Haque}}, \bibinfo {author} {\bibfnamefont {S.}~\bibnamefont {Kee}}, \bibinfo
  {author} {\bibfnamefont {D.~R.}\ \bibnamefont {Villalva}}, \bibinfo {author}
  {\bibfnamefont {W.-L.}\ \bibnamefont {Ong}},\ and\ \bibinfo {author}
  {\bibfnamefont {D.}~\bibnamefont {Baran}},\ }\bibfield  {title} {\bibinfo
  {title} {Halide perovskites: {Thermal} transport and prospects for
  thermoelectricity},\ }\href {https://doi.org/10.1002/advs.201903389}
  {\bibfield  {journal} {\bibinfo  {journal} {Adv. Sci.}\ }\textbf {\bibinfo
  {volume} {7}},\ \bibinfo {pages} {1903389} (\bibinfo {year}
  {2020})}\BibitemShut {NoStop}%
\bibitem [{\citenamefont {Wang}\ \emph {et~al.}(2018)\citenamefont {Wang},
  \citenamefont {Lin}, \citenamefont {Zhu}, \citenamefont {Zheng},
  \citenamefont {Wang}, \citenamefont {Li},\ and\ \citenamefont
  {Zhu}}]{wang2018cation}%
  \BibitemOpen
  \bibfield  {author} {\bibinfo {author} {\bibfnamefont {Y.}~\bibnamefont
  {Wang}}, \bibinfo {author} {\bibfnamefont {R.}~\bibnamefont {Lin}}, \bibinfo
  {author} {\bibfnamefont {P.}~\bibnamefont {Zhu}}, \bibinfo {author}
  {\bibfnamefont {Q.}~\bibnamefont {Zheng}}, \bibinfo {author} {\bibfnamefont
  {Q.}~\bibnamefont {Wang}}, \bibinfo {author} {\bibfnamefont {D.}~\bibnamefont
  {Li}},\ and\ \bibinfo {author} {\bibfnamefont {J.}~\bibnamefont {Zhu}},\
  }\bibfield  {title} {\bibinfo {title} {Cation dynamics governed thermal
  properties of lead halide perovskite nanowires},\ }\href
  {https://doi.org/10.1021/acs.nanolett.7b04437} {\bibfield  {journal}
  {\bibinfo  {journal} {Nano Lett.}\ }\textbf {\bibinfo {volume} {18}},\
  \bibinfo {pages} {2772} (\bibinfo {year} {2018})}\BibitemShut {NoStop}%
\bibitem [{\citenamefont {Seshadri}\ \emph {et~al.}(1999)\citenamefont
  {Seshadri}, \citenamefont {Baldinozzi}, \citenamefont {Felser},\ and\
  \citenamefont {Tremel}}]{seshadri1999visualizing}%
  \BibitemOpen
  \bibfield  {author} {\bibinfo {author} {\bibfnamefont {R.}~\bibnamefont
  {Seshadri}}, \bibinfo {author} {\bibfnamefont {G.}~\bibnamefont
  {Baldinozzi}}, \bibinfo {author} {\bibfnamefont {C.}~\bibnamefont {Felser}},\
  and\ \bibinfo {author} {\bibfnamefont {W.}~\bibnamefont {Tremel}},\
  }\bibfield  {title} {\bibinfo {title} {Visualizing electronic structure
  changes across an antiferroelectric phase transition: {Pb$_2$MgWO$_6$}},\
  }\href {https://doi.org/10.1039/a904408f} {\bibfield  {journal} {\bibinfo
  {journal} {J. Mater. Chem.}\ }\textbf {\bibinfo {volume} {9}},\ \bibinfo
  {pages} {2463} (\bibinfo {year} {1999})}\BibitemShut {NoStop}%
\bibitem [{\citenamefont {Raulot}\ \emph {et~al.}(2002)\citenamefont {Raulot},
  \citenamefont {Baldinozzi}, \citenamefont {Seshadri},\ and\ \citenamefont
  {Cortona}}]{raulot2002ab}%
  \BibitemOpen
  \bibfield  {author} {\bibinfo {author} {\bibfnamefont {J.-M.}\ \bibnamefont
  {Raulot}}, \bibinfo {author} {\bibfnamefont {G.}~\bibnamefont {Baldinozzi}},
  \bibinfo {author} {\bibfnamefont {R.}~\bibnamefont {Seshadri}},\ and\
  \bibinfo {author} {\bibfnamefont {P.}~\bibnamefont {Cortona}},\ }\bibfield
  {title} {\bibinfo {title} {An ab-initio study of the r\^ole of lone pairs in
  the structure and insulator{\textendash}metal transition in {SnO} and
  {PbO}},\ }\href {https://doi.org/10.1016/s1293-2558(02)01280-3} {\bibfield
  {journal} {\bibinfo  {journal} {Solid State Sci.}\ }\textbf {\bibinfo
  {volume} {4}},\ \bibinfo {pages} {467} (\bibinfo {year} {2002})}\BibitemShut
  {NoStop}%
\bibitem [{\citenamefont {Stoltzfus}\ \emph {et~al.}(2007)\citenamefont
  {Stoltzfus}, \citenamefont {Woodward}, \citenamefont {Seshadri},
  \citenamefont {Klepeis},\ and\ \citenamefont
  {Bursten}}]{stoltzfus2007structure}%
  \BibitemOpen
  \bibfield  {author} {\bibinfo {author} {\bibfnamefont {M.~W.}\ \bibnamefont
  {Stoltzfus}}, \bibinfo {author} {\bibfnamefont {P.~M.}\ \bibnamefont
  {Woodward}}, \bibinfo {author} {\bibfnamefont {R.}~\bibnamefont {Seshadri}},
  \bibinfo {author} {\bibfnamefont {J.-H.}\ \bibnamefont {Klepeis}},\ and\
  \bibinfo {author} {\bibfnamefont {B.}~\bibnamefont {Bursten}},\ }\bibfield
  {title} {\bibinfo {title} {Structure and bonding in {SnWO$_4$,} {PbWO$_4$,}
  and {BiVO$_4$:} lone pairs vs inert pairs},\ }\href
  {https://doi.org/10.1021/ic061157g} {\bibfield  {journal} {\bibinfo
  {journal} {Inorg. Chem.}\ }\textbf {\bibinfo {volume} {46}},\ \bibinfo
  {pages} {3839} (\bibinfo {year} {2007})}\BibitemShut {NoStop}%
\bibitem [{\citenamefont {Hellman}\ \emph {et~al.}(2013)\citenamefont
  {Hellman}, \citenamefont {Steneteg}, \citenamefont {Abrikosov},\ and\
  \citenamefont {Simak}}]{hellman2013temperature}%
  \BibitemOpen
  \bibfield  {author} {\bibinfo {author} {\bibfnamefont {O.}~\bibnamefont
  {Hellman}}, \bibinfo {author} {\bibfnamefont {P.}~\bibnamefont {Steneteg}},
  \bibinfo {author} {\bibfnamefont {I.~A.}\ \bibnamefont {Abrikosov}},\ and\
  \bibinfo {author} {\bibfnamefont {S.~I.}\ \bibnamefont {Simak}},\ }\bibfield
  {title} {\bibinfo {title} {Temperature dependent effective potential method
  for accurate free energy calculations of solids},\ }\href@noop {} {\bibfield
  {journal} {\bibinfo  {journal} {Physical Review B}\ }\textbf {\bibinfo
  {volume} {87}},\ \bibinfo {pages} {104111} (\bibinfo {year}
  {2013})}\BibitemShut {NoStop}%
\bibitem [{\citenamefont {Li}\ \emph {et~al.}(2017)\citenamefont {Li},
  \citenamefont {Jin}, \citenamefont {Cui}, \citenamefont {Zhuang},
  \citenamefont {Zhang}, \citenamefont {Meng}, \citenamefont {Bao},
  \citenamefont {Liu},\ and\ \citenamefont {Zhou}}]{li2017unexpected}%
  \BibitemOpen
  \bibfield  {author} {\bibinfo {author} {\bibfnamefont {Y.}~\bibnamefont
  {Li}}, \bibinfo {author} {\bibfnamefont {X.}~\bibnamefont {Jin}}, \bibinfo
  {author} {\bibfnamefont {T.}~\bibnamefont {Cui}}, \bibinfo {author}
  {\bibfnamefont {Q.}~\bibnamefont {Zhuang}}, \bibinfo {author} {\bibfnamefont
  {D.}~\bibnamefont {Zhang}}, \bibinfo {author} {\bibfnamefont
  {X.}~\bibnamefont {Meng}}, \bibinfo {author} {\bibfnamefont {K.}~\bibnamefont
  {Bao}}, \bibinfo {author} {\bibfnamefont {B.}~\bibnamefont {Liu}},\ and\
  \bibinfo {author} {\bibfnamefont {Q.}~\bibnamefont {Zhou}},\ }\bibfield
  {title} {\bibinfo {title} {Unexpected stable stoichiometries and
  superconductivity of potassium-rich sulfides},\ }\href@noop {} {\bibfield
  {journal} {\bibinfo  {journal} {RSC Advances}\ }\textbf {\bibinfo {volume}
  {7}},\ \bibinfo {pages} {44884} (\bibinfo {year} {2017})}\BibitemShut
  {NoStop}%
\bibitem [{\citenamefont {Sangster}\ and\ \citenamefont
  {Pelton}(1997)}]{sangster1997ks}%
  \BibitemOpen
  \bibfield  {author} {\bibinfo {author} {\bibfnamefont {J.}~\bibnamefont
  {Sangster}}\ and\ \bibinfo {author} {\bibfnamefont {A.}~\bibnamefont
  {Pelton}},\ }\bibfield  {title} {\bibinfo {title} {The {KS}
  (potassium-sulfur) system},\ }\href@noop {} {\bibfield  {journal} {\bibinfo
  {journal} {Journal of Phase Equilibria}\ }\textbf {\bibinfo {volume} {18}},\
  \bibinfo {pages} {82} (\bibinfo {year} {1997})}\BibitemShut {NoStop}%
\bibitem [{\citenamefont {B{\"o}ttcher}\ \emph {et~al.}(1993)\citenamefont
  {B{\"o}ttcher}, \citenamefont {Getzschmann},\ and\ \citenamefont
  {Keller}}]{bottcher1993kenntnis}%
  \BibitemOpen
  \bibfield  {author} {\bibinfo {author} {\bibfnamefont {P.}~\bibnamefont
  {B{\"o}ttcher}}, \bibinfo {author} {\bibfnamefont {J.}~\bibnamefont
  {Getzschmann}},\ and\ \bibinfo {author} {\bibfnamefont {R.}~\bibnamefont
  {Keller}},\ }\bibfield  {title} {\bibinfo {title} {Zur kenntnis der
  dialkalimetalldichalkogenide {$\beta$-Na$_2$S$_2$}, {K$_2$S$_2$},
  {$\alpha$-Rb$_2$S$_2$}, {$\beta$-Rb$_2$S$_2$}, {K$_2$Se$_2$}, {Rb$_2$Se$_2$},
  {$\alpha$-K$_2$Te$_2$}, {$\beta$-K$_2$Te$_2$} und {Rb$_2$Te$_2$}},\
  }\href@noop {} {\bibfield  {journal} {\bibinfo  {journal} {Zeitschrift
  f{\"u}r anorganische und allgemeine Chemie}\ }\textbf {\bibinfo {volume}
  {619}},\ \bibinfo {pages} {476} (\bibinfo {year} {1993})}\BibitemShut
  {NoStop}%
\end{thebibliography}

\end{document}